\documentclass[12pt]{article}

\usepackage{authordate1-4}
\usepackage{fullpage}
\usepackage{indentfirst}
\usepackage{latexsym}
\usepackage{amsmath}
\usepackage{amsthm}
\usepackage{amssymb}
\usepackage{times}
\usepackage{graphicx}
\usepackage{algorithmic}
\usepackage[ruled]{algorithm}

\DeclareMathOperator*{\mycup}{\cup}

\DeclareMathOperator{\mytimes}{\times}

\renewcommand{\Pr}{\mathsf{Pr}}
\newcommand{\E}{\mathsf{E}}
\newcommand{\V}{\mathsf{Var}}

\newcommand{\naturals}{\mathbb{N}}

\newcommand{\normal}{\mathsf{N}}

\newcommand{\Uni}{\mathsf{Uni}}

\begin{document}

\title{Dynamic Markov Bases}

\author{Adrian Dobra\footnote{Adrian Dobra is Assistant Professor, Departments of Statistics, Biobehavioral Nursing and Health Systems and the Center for Statistics and the Social Sciences, University of Washington, Seattle, WA 98195-4322 (email: adobra@uw.edu).}}

\maketitle

\begin{abstract}
 We present a computational approach for generating Markov bases for multi-way contingency tables whose cells counts might be constrained by fixed marginals and by lower and upper bounds. Our framework includes tables with structural zeros as a particular case. Instead of computing the entire Markov basis in an initial step, our framework finds sets of local moves that connect each table in the reference set with a set of neighbor tables. We construct a Markov chain on the reference set of tables that requires only a set of local moves at each iteration. The union of these sets of local moves forms a dynamic Markov basis. We illustrate the practicality of our algorithms in the estimation of exact p-values for a three-way table with structural zeros and a sparse eight-way table. Computer code implementing the methods described in the article as well as the two datasets used in the numerical examples are available as supplemental material.
\end{abstract}

\section{Introduction} \label{sec:intro}

\noindent Sampling from sets of contingency tables is key for performing exact conditional tests. Such tests arise by eliminating nuisance parameters through conditioning on their minimal sufficient statistics \cite{agresti-1992}. They are needed when the validity of asymptotic approximations to the null distributions of test statistics of interest is questionable or when no such approximations are available. Kreiner \shortcite{kreiner-1987} argues against the use of large-sample $\chi^2$ approximations for goodness-of-fit tests for large sparse tables, while Haberman \shortcite{haberman-1988} raises similar concerns for tables having expected cell counts that are small and large. The problem is further compounded by the existence of structural zeros or by limits on the values allowed on each cell, e.g. occurrence matrices in ecological studies \cite{chen-jasa2005}.\\
\indent One of the earlier algorithms for sampling two-way contingency tables with fixed row and column totals is due to Mehta and Patel \shortcite{mehta-patel-1983}. Other key developments include the importance sampling approaches of Booth and Butler \shortcite{booth-butler-1999}, Chen et al. \shortcite{chen-jasa2005}, Chen et al. \shortcite{chen-annals2006} and Dinwoodie and Chen \shortcite{dinwoodie-chen-2010}. Various Markov chain algorithms have been proposed by Besag and Clifford \shortcite{besag-clifford-1989}, Guo and Thompson \shortcite{guo-thompson-1992}, Forster et al. \shortcite{forster-et-1996} and Caffo and Booth \shortcite{caffo-booth-2001}. A very good review is presented in Caffo and Booth \shortcite{caffo-booth-2003}.\\
\indent One of the central contributions to the literature was the seminal paper by Diaconis and Sturmfels \shortcite{diaconis-sturmfels-1998}. They generate tables in a reference set $T$ through a Markov basis. The fundamental concept behind a Markov basis is easily understood by considering all the possible pairwise differences of tables in $T$, i.e. $\mathcal{M} = \{ n^{\prime}-n^{\prime\prime}:n^{\prime},n^{\prime\prime}\in T\}$. The elements of $\mathcal{M}$ are called moves. Any table $n^{\prime}\in T$ can be transformed in another table $n^{\prime\prime}\in T$ by applying the move $n^{\prime\prime}-n^{\prime}\in \mathcal{M}$. Clearly not all the moves in $\mathcal{M}$ are needed to connect any two tables in $T$ through a series of moves. A Markov basis for $T$ is obtained by eliminating some of the moves in $\mathcal{M}$ such that the remaining moves still connect $T$. Generating a Markov basis is in the most general case a computationally difficult task that is solved using computational algebraic techniques. The simplest Markov basis contains only moves with two entries equal to $1$, two entries equal to $-1$ and the remaining entries equal to zero. It connects all the two-way tables with fixed row and columns totals \cite{diaconis-sturmfels-1998}. These primitive moves extend to decomposable log-linear models as described in Dobra \shortcite{dobra-2003}. A divide-and-conquer technique for the determination of Markov bases for reducible log-linear models is given in Dobra and Sullivant \shortcite{dobra-sullivant-2004}. Additional information on Markov bases can be found in Drton et al. \shortcite{drton-sturmfels-sullivant-2009}.\\
\indent In this paper we focus on the general problem of the determination of a Markov basis for sets of multi-way tables defined by fixed marginals and by lower and upper bounds constraints on each cell count. Bounds constraints arise in disclosure limitation from information deemed to be public at a certain time \cite{willenborg-dewaal-2000}. In ecological inference lower bounds constrains are induced by individual-level information \cite{wakefield-2004}. Noteworthy theoretical contributions on Markov bases for bounded tables include Aoki and Takemura \shortcite{aoki-takemura-2005}, Rapallo \shortcite{rapallo-2006}, Rapallo and Rogantin \shortcite{rapallo-rogantin-2007}, Aoki and Takemura \shortcite{aoki-takemura-2010}, and Rapallo and Yoshida \shortcite{rapallo-yoshida-2010}. Unfortunately, it is quite difficult to carry out a principled assessment of the practical value of their algebraic statistics results for tables with more than two dimensions due to the absence of dedicated software that would make these methods accessible to lay users.\\
\indent So far the papers dedicated to Markov bases have attempted to generate them in a preliminary step that needs to be completed before the corresponding random walk can be started. In practice this step can be computationally prohibitive to perform because the resulting Markov bases contain a very large number of elements even for three-way tables \cite{deloera-onn-2005}. The Markov bases repository of Kahle and Rauh (\texttt{http://mbdb.mis.mpg.de}) is very useful for understanding the complexity of the moves even for simple, non-decomposable log-linear models. We avoid this major computational hurdle by developing dynamic Markov bases. Such bases do not have to be generated in advance. Instead, at each iteration of our Markov chain algorithm we sample from a set of local moves that connect the table that represents the current state of the chain with a set of neighbor tables. Our computational approach extends the applicability of Markov bases to examples that could not be handled with other approaches presented in the literature.\\
\indent The structure of the paper is as follows. In Section \ref{sec:framework} we present the notations and the  setting of our framework. In Section \ref{sec:dmb} we introduce dynamic Markov bases and present two algorithms for sampling multi-way tables. In Section \ref{sec:art} we discuss these algorithms in the context of the importance sampling approaches of Booth and Butler \shortcite{booth-butler-1999} and Chen et al. \shortcite{chen-annals2006}. In Sections \ref{sec:markovchain} and \ref{sec:finding} we give our Markov chain algorithm based on dynamic Markov bases. In Section \ref{sec:examples} we illustrate the applicability of our methodology for a three-way table with structural zeros and a sparse eight-way table. In Section \ref{sec:conclusions} we make concluding remarks.

\section{Notations and Framework} \label{sec:framework}

\noindent Let $X=(X_{1},X_{2},\ldots,X_{k})$ be a vector of discrete random variables. Variable $X_{j}$ takes values $x_{j}\in \mathcal{I}_j=\{1,2,\ldots,I_j\}$, $I_j\ge 2$. Consider a contingency table $n=\{n(i)\}_{i\in \mathcal{I}}$ of observed counts associated with $X$, where $\mathcal{I}=\mytimes\limits_{j=1}^{k}I_{j}$ are cell indices. The set $\mathcal{I}$ is assumed to be ordered lexicographically, so that $\mathcal{I}=\left\{ i^{1},i^{2},\ldots,i^{m_{c}}\right\}$, where $i^{1}=(1,\ldots,1,1)$, $i^{2}=(1,\ldots,1,2)$, $i^{m_{c}}=(I_{1},I_{2},\ldots,I_{k})$ and $m_{c}=I_{1}\cdot I_{2}\cdot\ldots\cdot I_{k}$ is the total number of cells. With this ordering the $k$-dimensional array $n=\{n(i)\}_{i\in \mathcal{I}}$ is written as a vector $n=\left\{ n(i^{1}),n(i^{2}),\ldots,n(i^{m_{c}})\right\}$. For $C\subset K=\{1,\ldots,k\}$, the $C$-marginal $n_{C}=\{ n_C(i_C)\}_{i_C\in \mathcal{I}_C}$ of $n$ is the cross-classification associated with the sub-vector $X_{C}$ of $X$, where $\mathcal{I}_C=\mytimes_{j\in C}\mathcal{I}_j$. The grand total of $n$ is $n_{\emptyset}$.\\
\indent Consider two other $k$-way tables $n^L$ and $n^U$ that define lower and upper bounds for $n$. The role of these bounds is to specify various constraints that might exist for the cell entries of $n$. For example, a structural zero in cell $i\in \mathcal{I}$ is specified as $n^L(i)=n^U(i)=0$. Zero-one tables are expressed by taking $n^L(i)=0$ and $n^U(i)=1$ for all $i\in \mathcal{I}$. In addition to the bounds constraints, the cell entries of $n$ can be required to satisfy a set of linear constraints induced by a set of fixed marginals $\{ n_{C}:C\in \mathcal{C}\}$, where $\mathcal{C}=\{ C_1,\ldots,C_q\}$, with $C_j\subset K$ for $j=1,\ldots,q$. We let $\mathcal{A}$ be a log-linear model whose minimal sufficient statistics are  $\{ n_{C}:C\in \mathcal{C}\}$. We define the set of tables that are consistent with the minimal sufficient statistics of $\mathcal{A}$ and with the bounds $n^L$ and $n^U$:
\begin{eqnarray} \label{eq:T}
 T=\left\{ n^{\prime}=\{n^{\prime}(i)\}_{i\in \mathcal{I}}: n^{\prime}_{C_j}=n_{C_j},\mbox{ for }j=1,\ldots,q, n^L(i)\le n'(i)\le n^U(i),\mbox{ for }i\in \mathcal{I}\right\}.
\end{eqnarray}
\noindent We assume that $n\in T$, that is, the bounds constraints $n^L$ and $n^U$ are not at odds with the observed data. The set $T$ induces bounds constraints $L(i)$ and $U(i)$ on each cell entry $n(i)$, $i\in \mathcal{I}$:
\begin{eqnarray*}
 L(i) = \min\left\{ n^{\prime}(i): n^{\prime}\in T\right\},\quad U(i) = \max\left\{ n^{\prime}(i): n^{\prime}\in T\right\}.
\end{eqnarray*} 
\noindent These bounds are possibly tighter than the initial bounds $n^L$ and $n^U$, i.e. 
$$ 
 0\le n^{L}(i)\le L(i)\le n^{\prime}(i)\le U(i)\le n^{U}(i)\le n_{\emptyset},
$$
\noindent for $i\in \mathcal{I}$ and $n^{\prime}\in T$. They can be determined by integer programming algorithms \cite{boyd-vandenberghe-2004} or by other methods such as the generalized shuttle algorithm \cite{dobra-fienberg-2008}. The constraints that define $T$ can lead to the exact determination of some cell counts. More explicitly, we consider $\mathcal{S}\subset \mathcal{I}$ to be the set of cells such that $L(i)<U(i)$. This means that all the tables in $T$ have the same counts for the cells in $\mathcal{I}\setminus \mathcal{S}$. We note that the determination of $\mathcal{S}$ needs to be made based on the bounds $L=\left\{L(i)\right\}_{i\in \mathcal{I}}$ and $U=\left\{U(i)\right\}_{i\in \mathcal{I}}$ and not on  $n^{L}$ and $n^{U}$. Thus the set $T$ comprises all the integer arrays $n^{\prime}$ that satisfy the equality constraints
\begin{eqnarray}\label{eq:linearconstraintsT}
n^{\prime}_{C_{j}}\left( i_{C_{j}}\right) & = & n_{C_{j}}\left( i_{C_{j}}\right), \mbox{ for }i_{C_{j}}\in \mathcal{I}_{C_{j}}, j=1,2,\ldots,q,\\
n^{\prime}(i) & = & n(i),\mbox{ for } i\in \mathcal{I}\setminus\mathcal{S},\nonumber
\end{eqnarray}
\noindent as well as the bounds constraints
\begin{eqnarray}\label{eq:boundsconstraintsT}
 & L(i)\le n^{\prime}(i)\le U(i),\; \mbox{ for } i\in \mathcal{S}.& 
\end{eqnarray}
By ordering the cell indices $\mathcal{I}$ lexicographically the equality constraints (\ref{eq:linearconstraintsT}) can be written as a linear system of equations
\begin{eqnarray} \label{eq:linsysaxb}
 A n^{\prime} & = & b,
\end{eqnarray}
\noindent where $A$ is a $m_{r}\mytimes m_{c}$ matrix with elements equal to $0$ or $1$, $m_{r}=\sum\limits_{j=1}^{q}\mid\mathcal{I}_{C_{j}}|+|\mathcal{I}\setminus\mathcal{S}|$ and $b=An$ is a $m_{r}$-dimensional column vector. Here $|E|$ denotes the number of elements of a set $E$. In order to simplify the notations we subsequently assume that $\mathcal{S}=\mathcal{I}$ with the understanding that the determination of $\mathcal{S}$ is key and needs to be completed before our algorithms are applied.\\
\indent  Two distributions defined on $T$ play a key role in statistical analyses. They are the uniform and the hypergeometric distributions
\begin{eqnarray} \label{eq:targetdistrib}
 P_U(n^{\prime}) =  \frac{1}{|T|}, \mbox { and }
 P_H(n^{\prime})  = \frac{\left[ \prod\limits_{i\in \mathcal{I}} n^{\prime}(i)!\right]^{-1}}{\sum\limits_{n^{\prime\prime}\in T}\left[ \prod\limits_{i\in \mathcal{I}} n^{\prime\prime}(i)!\right]^{-1}},
\end{eqnarray}
\noindent for $n'\in T$. In the most general case, the normalizing constants of $P_{H}(\cdot)$ and $P_U(\cdot)$ can be computed only if $T$ can be enumerated. Sundberg \shortcite{sundberg-1975} developed a formula for the normalizing constant of $P_H(\cdot)$ if $\mathcal{A}$ is decomposable and there are no bounds constraints (i.e. $n^L(i)=0$ and $n^U(i)=n_{\emptyset}$ for all $i\in \mathcal{I}$). Sampling from $P_U(\cdot)$ is relevant for estimating the number of tables in $T$ \cite{chen-jasa2005,dobra-jspi2006} or for performing the conditional volume test  \cite{diaconis-efron-1985}. The hypergeometric distribution $P_H(\cdot)$ arises by conditioning on the log-linear model $\mathcal{A}$ and the set of tables $T$ under multinomial sampling. Haberman \shortcite{haberman-1974} proved that the the log-linear interaction terms cancel out, which leads to equation (\ref{eq:targetdistrib}).\\
\indent Sampling from $P_U(\cdot)$ and $P_H(\cdot)$ is straightforward if $T$ can be explicitly determined, but this task is computationally infeasible for most real-world datasets. The goal of this paper is to develop a sampling procedure from $P_H(\cdot)$ and $P_U(\cdot)$ for any set of tables $T$ induced by a set of fixed marginals and lower and upper bounds arrays.

\section{Dynamic Markov Bases} \label{sec:dmb}

\noindent Producing an entire Markov basis up-front is computationally expensive; it also makes random walks impractical for reference sets $T$ involving sparse high-dimensional tables. Such bases contain an extremely large number of moves that are difficult to handle in the rare cases when they can actually be found using an algebra package.  However, one does not necessarily need to know the entire Markov basis in order to run a Markov chain on $T$. The Markov bases we introduce in this section are dynamic because they are not generated ahead of time. They consist of sets of moves that connect a given table $n^{*}\in T$ with a set of neighbor tables $\mbox{nbd}_{T}(n^{*})\subseteq T$. The union of the sets of neighbor tables should be symmetric (i.e., $n^{\prime}\in \mbox{nbd}_{T}(n^{\prime\prime})$ if and only if $n^{\prime\prime}\in \mbox{nbd}_{T}(n^{\prime})$), and their union should connect $T$, i.e., 
\begin{eqnarray}\label{eq:localbasis}
\bigcup_{n^{\prime}\in T} \{ n^{\prime\prime}-n^{\prime}:n^{\prime\prime}\in \mbox{nbd}_{T}(n^{\prime})\}
\end{eqnarray}
\noindent is a Markov basis for $T$. The moves given by the difference between a table $n^{\prime}$ and one of its neighbors $n^{\prime\prime}\in  \mbox{nbd}_{T}(n^{\prime})$ are called local.\\
\indent  The sets of neighbors are determined as follows. For two integers $a\le b$, we denote $(a:b)=\{a,a+1,\ldots,b\}$. We define $(a:b)=\emptyset$ if $a>b$. Let $\Delta_{m_{c}}$ denote the set of all permutations of $(1:m_{c})$. For a permutation $\delta\in \Delta_{m_{c}}$, we define the set  of tables $T_{\delta}$ that is obtained by reordering the cell counts of tables in $T$ according to $\delta$. The re-ordered version $n^{*}_{\delta}\in T_{\delta}$ of $n^{*}$ is such that $n^{*}_{\delta}(i^{j}) = n^{*}(i^{\delta(j)})$ for $1\le j\le m_{c}$. The difference between $T_{\delta}$ and $T$ relates to the ordering of their cells. We have $T=T_{\delta_{0}}$ where $\delta_{0}\in \Delta_{m_{c}}$, $\delta_{0}(j)=j$ for $1\le j\le m_{c}$.\\
\indent For a table $n^{*}\in T$ and an index $s\in (1:m_{c})$, we define the set of tables that have the same counts in cells $\{i^{\delta(1)},\ldots,i^{\delta(s)}\}$ as table $n^{*}$:
\begin{eqnarray}\label{eq:tdelta}
 T_{\delta,s}(n^{*}_{\delta}) = \left\{ n^{\prime}_{\delta}\in T_{\delta}:n^{\prime}_{\delta}\left(i^{j}\right)=n^{*}_{\delta}\left(i^{j}\right), \mbox{ for }j=1,\ldots,s\right\}.
\end{eqnarray}
We define $T_{\delta,0}(n^{*}_{\delta}) = T_{\delta}$. We have $T_{\delta,m_{c}}(n^{*}_{\delta}) = \{n^{*}_{\delta}\}$,   and $n^{*}_{\delta}\in T_{\delta,s}(n^{*}_{\delta})$ for any $s\in (0:m_{c})$. The sets of tables $T_{\delta,s}(n^{*}_{\delta})$ become smaller as the number of common cells increases, i.e. $T_{\delta,s}(n^{*}_{\delta}) \supseteq T_{\delta,s'}(n^{*}_{\delta})$ for $0\le s \le s'\le m_{c}$. We consider the minimum and the maximum values of cell $i^{j}$ in the set of tables $T_{\delta,s}(n^{*}_{\delta})$, i.e.
\begin{eqnarray*}
 L_{\delta,n^{*},s}(i^{j}) = \min\left\{ n^{\prime}_{\delta}(i^{j}):n^{\prime}_{\delta}\in T_{\delta,s}(n^{*}_{\delta}) \right\},\quad U_{\delta,n^{*},s}(i^{j})  = \max\left\{ n^{\prime}_{\delta}(i^{j}):n^{\prime}_{\delta}\in T_{\delta,s}(n^{*}_{\delta}) \right\}.
\end{eqnarray*}
Remark that $L_{\delta,n^{*},s}(i^{j}) =U_{\delta,n^{*},s}(i^{j})=n^{*}_{\delta}(i^{j})$ for $j\in (1:s)$. Determining the minimum and maximum values for the remaining cells without exhaustively enumerating $T_{\delta,s}(n^{*}_{\delta})$ can be done by computing the integer lower and upper bounds induced on each cell by the constraints that define this set of tables. For $j\in ((s+1):m_{c})$,  $L_{\delta,n^{*},s}(i^{j})$ and $U_{\delta,n^{*},s}(i^{j})$ are the solutions of the linear programming problems
\begin{eqnarray} \label{eq:linprogbounds}
 \mbox{minimize} & \pm n^{\prime}_{\delta}(i^{j}) &\\
 \mbox{subject to} & An^{\prime} = b, &\nonumber \\ 
 & L(i)\le n^{\prime}(i)\le U(i), & \mbox{for } i\in \mathcal{I},\nonumber \\
 & n^{\prime}_{\delta}\left(i^{j}\right)=n^{*}_{\delta}\left(i^{j}\right), & \mbox{for } j=1,\ldots,s,\nonumber\\
 & n^{\prime}(i) \in  \naturals, & \mbox{for } i\in \mathcal{I}.\nonumber 
\end{eqnarray}
Here $\naturals$ is the set of nonnegative integers. Computationally it is quite demanding to determine the integer bounds $L_{\delta,n^{*},s}(i^{j})$ and $U_{\delta,n^{*},s}(i^{j})$, hence we approximate them with the integer counterparts of the real bounds $L^{R}_{\delta,n^{*},s}(i^{j})$ and $U^{R}_{\delta,n^{*},s}(i^{j})$. These real bounds are calculated by solving the optimization problems (\ref{eq:linprogbounds}) without the constraints $n^{\prime}(i) \in  \naturals$, for $i\in \mathcal{I}$. In general, we have
$$
 L_{\delta,n^{*},s}(i^{j})  \ge \left\lceil L^{R}_{\delta,n^{*},s}(i^{j})\right\rceil, \quad U_{\delta,n^{*},s}(i^{j}) \le \left\lfloor U^{R}_{\delta,n^{*},s}(i^{j})\right\rfloor.
$$
We denote by $\lceil a\rceil$ and $\lfloor a\rfloor$ the smallest integer greater than or equal to $a$ and the largest integer smaller than or equal to $a$, respectively. For the purpose of implementing the procedures described in this paper the approximation given by rounding the real bounds seems to perform well.\\
\indent We describe a method for randomly sampling a table in $T_{\delta}$. Algorithm \ref{alg:randomtable} generates a feasible table by sequentially sampling the count of each cell given that the counts of the cells preceding it in the reordering of $\mathcal{I}$ defined by $\delta$ have already been fixed. The permutation $\delta$ defines the order in which the cell counts are sampled. The set of possible values of each cell are defined by the lower and upper bounds induced by the constraints that define $T$ and the cell counts already determined. This procedure is employed at each iteration of the sequential importance sampling (SIS) algorithm \cite{chen-annals2006,dinwoodie-chen-2010} and has also been suggested, in various forms, in other papers \cite{chen-jasa2005,dobra-jspi2006,chen-jcgs2007}. We note that the determination of multi-way tables through a sequential adjustment of cell bounds appears in earlier writings such as Dobra \shortcite{dobra-2002} who proposes a branch-and-bound algorithm for enumerating all the multi-way tables consistent with a set of linear and bounds constraints, as well as Dobra et al. \shortcite{dobra-karr-sanil-2003} and Dobra and Fienberg \shortcite{dobra-fienberg-2008} who develop the generalized shuttle algorithm.

\begin{algorithm}
\caption{Sample a table $n^{\prime}_{\delta}\in T_{\delta}$}
\label{alg:randomtable}
\begin{algorithmic}[1]
\STATE Consider a table $n^{\prime}_{\delta}$ whose cells are currently unoccupied.
\STATE Set $s\leftarrow 1$.
\WHILE{$s\le m_{c}$}
 \STATE Calculate the updated bounds for cell $i^{s}$. If $s=1$, set $L^{\prime}_{\delta,n^{\prime},0}(i^{1})=L(i^{\delta(1)})$ and $U^{\prime}_{\delta,n^{\prime},0}(i^{1})=U(i^{\delta(1)})$. Otherwise solve the linear programming problems (\ref{eq:linprogbounds}) to determine the real bounds for cell $i^{s}$ and set $L^{\prime}_{\delta,n^{\prime},s-1}(i^{s})=\left \lceil L^{R}_{\delta,n^{\prime},s-1}(i^{s})\right\rceil$ and $U^{\prime}_{\delta,n^{\prime},s-1}(i^{s})=\left\lfloor U^{R}_{\delta,n^{\prime},s-1}(i^{s})\right\rfloor$.
 \IF{$L^{\prime}_{\delta,n^{\prime},s-1}(i^{s})>U^{\prime}_{\delta,n^{\prime},s-1}(i^{s})$} 
 \STATE STOP.  \COMMENT{The algorithm terminates without generating any table.}
\ELSE
	\IF{$L^{\prime}_{\delta,n^{\prime},s-1}(i^{s})=U^{\prime}_{\delta,n^{\prime},s-1}(i^{s})$}
		\STATE Set $n^{\prime}_{\delta}(i^{s})\leftarrow L^{\prime}_{\delta,n^{\prime},s-1}(i^{s})$.
	\ELSE
	\STATE Sample a cell value $n^{\prime}_{\delta}(i^{s})$ from a discrete distribution $f^{\left(L^{\prime}_{\delta,n^{\prime},s-1}(i^{s}):U^{\prime}_{\delta,n^{\prime},s-1}(i^{s})\right)}(\cdot)$ with support $\left( L^{\prime}_{\delta,n^{\prime},s-1}(i^{s}):U^{\prime}_{\delta,n^{\prime},s-1}(i^{s})\right)$.
	\ENDIF
	\STATE Go to the next cell by setting $s\leftarrow s+1$.
\ENDIF
\ENDWHILE
\RETURN{$n^{\prime}_{\delta}$}
\end{algorithmic}
\end{algorithm}

Algorithm \ref{alg:randomtable} ends at line 6 without returning a table if the combination of cell values chosen at the previous iterations does not correspond with any table in $T_{\delta}$. Such combinations could arise because there are gaps between the bounds that correspond with integers for which there do not exist any tables in $T$ associated with them. This issue has been properly recognized and discussed in Chen et al. \shortcite{chen-annals2006} who also propose conditions which they call the sequential interval property that check whether gaps exist for certain tables and configurations of fixed marginals. To the best of the authors' knowledge, there are no computational tools that implement these conditions. Once such tools become available, Algorithm \ref{alg:randomtable} could be improved by replacing lines 5 and 6 with a procedure for identifying which integers in $\left( L^{\prime}_{\delta,n^{\prime},s-1}(i^{s}):U^{\prime}_{\delta,n^{\prime},s-1}(i^{s})\right)$ actually correspond to least one table in $T$. This set of integers becomes the support of the discrete distribution from line 11. With this refinement  Algorithm \ref{alg:randomtable} will always return a valid table. In the numerical examples from Section \ref{sec:examples} we use the reciprocal distribution $f_{r}^{L,U}(v) \propto 1/(1+v)$ to sample a cell value at line 11. Other possible choices include the uniform distribution $f_{u}^{L,U}(v) = 1/(U-L+1)$ or the hypergeometric distribution $f_{h}^{L,U}(v) = {U \choose v}{U \choose L+U-v}/{2U \choose L+U}$.\\
\indent Algorithm \ref{alg:randomtable} finds any table $n^{*}_{\delta}\in T_{\delta}$ with strictly positive probability
\begin{eqnarray}\label{eq:sampleneighbor}
 \pi_{\delta,f^{(L:U)}(\cdot)}(n^{*}_{\delta}) \propto \prod\limits_{s=1}^{m_{c}} f^{(L^{\prime}_{\delta,n^{*},s-1}(i^{s}):U^{\prime}_{\delta,n^{*},s-1}(i^{s}))}(n^{*}_{\delta}(i^{s})).
\end{eqnarray}
We define the neighbors of $n^{*}_{\delta}$ as the set of tables returned by Algorithm \ref{alg:randomtable}, i.e. $\mbox{nbd}_{T_{\delta}}(n^{*}_{\delta})=T_{\delta}$. The corresponding set of local moves (\ref{eq:localbasis}) is a Markov basis for $T_{\delta}$. Since $T$ and $T_{\delta}$ are in a one-to-one correspondence, this is also a Markov basis for $T$. This Markov basis is dynamic because its moves are sampled using Algorithm \ref{alg:randomtable} from the distribution (\ref{eq:sampleneighbor}).\\
\indent  Algorithm \ref{alg:randomtable} returns a table in $T_{\delta}$ only after it has computed lower and upper bounds for each cell in $\mathcal{I}$. Calculating $2m_{c}$ bounds to generate one feasible table could be quite expensive especially for high-dimensional sparse tables. The counts of zero that characterize such tables are likely to make quite a few cells take only one possible value given the current values of the cells that have been already fixed \--- see lines 8 and 9. Therefore the efficiency of Algorithm 1 can be increased by identifying these fixed-value cells without computing bounds. We consider an array $x=\{ x(i^{1}),\ldots,x(i^{m_{c}})\}$. We transform the linear system of equations (\ref{eq:linsysaxb}) defined by the equality constraints (\ref{eq:linearconstraintsT}) by reordering the columns $i^{1},\ldots,i^{m_{c}}$ of the matrix $A$ according to $\delta$. The reordered versions of $A$ and $x$ are $A_{\delta}$ and $x_{\delta}$. The column of $A_{\delta}$ that corresponds with $x_{\delta}(i^{j})$ is equal with the column of $A$ that corresponds with $x(i^{\delta(j)})$. An equivalent form of the linear system (\ref{eq:linsysaxb}) is
\begin{eqnarray} \label{eq:linsysdelta}
 A_{\delta}x_{\delta} = b.
\end{eqnarray}
We take the augmented $m_{r}\times (m_{c}+1)$ matrix $[ A_{\delta}\mid b]$ obtained by stacking $A_{\delta}$ and $b$ along side each other. We determine the reduced row echelon form (RREF) $[\widehat{A}_{\delta}\mid\widehat{b}]$ of $[ A_{\delta}\mid b]$ using Gauss-Jordan elimination with partial pivoting \--- see, for example, Shores \shortcite{shores-2007}. The linear system (\ref{eq:linsysdelta}) is equivalent with
\begin{eqnarray} \label{eq:rref}
 \widehat{A}_{\delta}x_{\delta} = \widehat{b},
\end{eqnarray}
whose number of rows $m_{r}^{\prime}\le \min\{m_{r},m_{c}\}$ is equal with the rank of $A$. Since the linear system (\ref{eq:rref}) has fewer equations than the initial linear system (\ref{eq:linsysaxb}), it is more efficient to make use of it when defining the linear programming problems (\ref{eq:linprogbounds}). A smaller number of constraints translates into reduced computing times in the determination of the bounds in line 4 of Algorithm \ref{alg:randomtable}. Furthermore it is possible to re-arrange the columns of $\widehat{A}_{\delta}$ and the coordinates of $x_{\delta}$ such that the system (\ref{eq:rref}) is written as
\begin{eqnarray} \label{eq:boundfree}
 I_{m_{r}^{\prime}} x_{\delta}^{B} + \widehat{A}_{\delta}^{R} x_{\delta}^{F} = \widehat{b},
\end{eqnarray}
where $x_{\delta}^{B}$ represent the $m_{r}^{\prime}$ bound variables of the equivalent linear systems (\ref{eq:linsysaxb}) and (\ref{eq:linsysdelta}), $x_{\delta}^{F}$ is the $(m_{c}-m_{r}^{\prime})$-dimensional vector of free variables and $I_{l}$ is the $l$-dimensional identity matrix. Once the values of the free cells $x_{\delta}^{F}$ are fixed, the values of the bound cells are immediately determined:
\begin{eqnarray} \label{eq:boundsfromfree}
 x_{\delta}^{B} = \widehat{b} - \widehat{A}_{\delta}^{R} x_{\delta}^{F}.
\end{eqnarray}

\begin{algorithm}
\caption{Sample a table $n^{\prime}_{\delta}\in T_{\delta}$ (RREF version)}
\label{alg:randomtablerref}
\begin{algorithmic}[1]
\STATE Consider a table $n^{\prime}_{\delta}$ whose cells are currently unoccupied.
\STATE Find the RREF of the linear system (\ref{eq:linsysdelta}). 
\STATE Sample values for the free cells $\{(n^{\prime}_{\delta})^{F}_{j}:1\le j\le (m_{c}-m^{\prime}_{r})\}$ as described in lines 4-13 of Algorithm \ref{alg:randomtable}.
\STATE Determine the values of the bound cells $(n^{\prime}_{\delta})^{B}$ using equation (\ref{eq:boundsfromfree}).
\IF {$(n^{\prime}_{\delta})^{B}$ does not contain negative entries}
\RETURN{ the table $n^{\prime}_{\delta}\in T_{\delta}$ determined by $(n^{\prime}_{\delta})^{F}$ and $(n^{\prime}_{\delta})^{B}$}
\ELSE
\STATE STOP. \COMMENT{The algorithm terminates without generating a table} 
\ENDIF
\end{algorithmic}
\end{algorithm}

\noindent This leads us to a new version Algorithm \ref{alg:randomtablerref} of Algorithm \ref{alg:randomtable}. The successful determination of a table in $T_{\delta}$ using Algorithm \ref{alg:randomtablerref} requires the calculation of $2(m_{c}-m^{\prime}_{r})$ bounds instead of $2m_{c}$ bounds as in Algorithm \ref{alg:randomtable}. Furthermore, the calculation of these bounds is faster because the reduced system (\ref{eq:rref}) is used. Lines 2 and 4 of Algorithm \ref{alg:randomtablerref} can be implemented efficiently using BLAS (Basic Linear Algebra Subprograms) Fortran routines for matrix manipulations, thus overall Algorithm \ref{alg:randomtablerref} has a significant computational gain over Algorithm \ref{alg:randomtable}. We point out that the determination of the RREF should be done for the system (\ref{eq:linsysdelta}) and not for the initial system (\ref{eq:linsysaxb}) since each permutation of cell indices could lead to different sets of bound and free cells. Empirically we observed that calculating the RREF is computationally inexpensive and can be efficiently performed at each application of Algorithm \ref{alg:randomtablerref}. Line 5 of Algorithm \ref{alg:randomtablerref} is needed because certain combinations of values for the free cells might not correspond to any table in $T$, in which case negative integers are found in one or several bound cells. When computing the lower bounds $L^{F}_{\delta,n^{\prime},j}$ and the upper bounds $U^{F}_{\delta,n^{\prime},j}$ for the $j$-th free cell $(n^{\prime}_{\delta})^{F}_{j}$ in line 3  of Algorithm \ref{alg:randomtablerref}, we add the linear constraints associated with the sampled values  of the first $(j-1)$ free cells and  make use of the reduced system (\ref{eq:rref}) in the corresponding linear programming problems (\ref{eq:linprogbounds}). The probability that  Algorithm \ref{alg:randomtablerref} samples a table $n^{\prime}_{\delta}\in T_{\delta}$ is strictly positive:
\begin{eqnarray}\label{eq:sampleneighborrref}
 \pi^{F}_{\delta,f^{(L:U)}(\cdot)}(n^{*}_{\delta}) \propto \prod\limits_{j=1}^{m_{c}-m^{\prime}_{r}} f^{(L^{F}_{\delta,n^{\prime},j}:U^{F}_{\delta,n^{\prime},j})}((n^{\prime}_{\delta})^{F}_{j}).
\end{eqnarray}


\section{State of the Art} \label{sec:art}

\noindent Algorithm \ref{alg:randomtable} is key for the sequential importance sampling (SIS) algorithm \cite{chen-annals2006,dinwoodie-chen-2010}. Tables from $T_{\delta}$ are sampled from the discrete distribution given in equation (\ref{eq:sampleneighbor}) and are further used to calculate importance sampling estimates of various quantities of interest. For example, when calculating exact p-values, tables sampled from the uniform and hypergeometric distributions $P_{U}(\cdot )$ and $P_{H}(\cdot )$ given in equation (\ref{eq:targetdistrib}) are needed, but cannot be obtained through a direct sampling procedure. Instead, tables sampled with Algorithm \ref{alg:randomtable} are obtained, but these tables yield reliable estimates of exact p-values only if the discrete distribution (\ref{eq:sampleneighbor}) is close to the target distributions $P_{U}(\cdot )$ or $P_{H}(\cdot )$. Various cells orderings $\delta\in \Delta_{m_{c}}$ and discrete distributions $f^{(L:U)}(\cdot)$ lead to discrete distributions (\ref{eq:sampleneighbor}) that could be quite far from a desired target distribution on $T$. Unfortunately there is no well defined computational procedure that allows the selection of $\delta\in \Delta_{m_{c}}$ and $f^{(L:U)}(\cdot)$ for any set of tables $T$ and any target distribution on $T$. The SIS algorithm as described by Chen et al. \shortcite{chen-jasa2005}, Chen et al. \shortcite{chen-annals2006}, Chen \shortcite{chen-jcgs2007} performs well for many applications, but completely fails for the two numerical examples we discuss in Section \ref{sec:examples} that involve a three-way table with structural zeros and a sparse eight-way binary table. In a recent contribution, Dinwoodie and Chen \shortcite{dinwoodie-chen-2010} propose a procedure for sequentially updating the discrete distribution $f^{(L:U)}(\cdot)$ from line 11 of  Algorithm \ref{alg:randomtable} as a function of the previously sampled cell values. With this improved version of SIS they obtain more promising results for the sparse eight-way binary table example. However, there is no theoretical argument which shows that the examination of other examples will not lead to situations in which SIS does not perform well due to the inability of Algorithm \ref{alg:randomtable} to sample tables that receive high probabilities under the target distribution. Replacing Algorithm \ref{alg:randomtable} with the more efficient Algorithm \ref{alg:randomtablerref} in an importance sampling procedure leads to improved computing times, but does not solve the critical issues related to finding appropriate choices of $\delta$ and $f^{(L:U)}(\cdot)$.\\
\indent Booth and Butler \shortcite{booth-butler-1999} proposed another approach for sampling multi-way tables. They start with a log-linear model $\mathcal{A}$ with minimal sufficient statistics $\left\{ n_{C}:C\in \mathcal{C}\right\}$ (see Section \ref{sec:framework}) and consider the expected cell values $\hat{\mu}=\left\{ \hat{\mu}(i^{1}),\ldots,\hat{\mu}(i^{m_{c}})\right\}$ under $\mathcal{A}$. Their sampling method is designed for a reference set of tables $T$ specified by the marginals $\left\{ n_{C}:C\in \mathcal{C}\right\}$. Since the ability to compute the expected cell values $\hat{\mu}$ is key, their framework does not extend to sets of tables that are also consistent with some lower and upper bounds $n^{L}$ and $n^{U}$. Therefore the sampling method of Booth and Butler \shortcite{booth-butler-1999} has a more limited domain of applicability than Algorithms \ref{alg:randomtable} and \ref{alg:randomtablerref}. Booth and Butler \shortcite{booth-butler-1999} consider a permutation of cell indices $\delta \in \Delta_{m_{c}}$ and partition the reordered cells $x_{\delta}$ as bound cells $x_{\delta}^{B}$ and free cells $x_{\delta}^{F}$ as in equation (\ref{eq:boundfree}). Furthermore, they assume that the cell counts follow independent normal distributions $x_{\delta}(i^{j})\sim \normal(\hat{\mu}(i^{j}),\hat{\mu}(i^{j}))$, $1\le j\le m_{c}$, which implies that the joint distribution of the free cells follows a multivariate normal distribution
\begin{eqnarray} \label{eq:mvnfree}
 x_{\delta}^{F} & \sim & \normal_{m_{c}-m^{\prime}_{r}}\left( \hat{\mu}_{\delta},\hat{V}^{\delta}\right),
\end{eqnarray}
\noindent where $\hat{V}^{\delta}=(\hat{v}^{\delta}_{j_{1}j_{2}})$ is a covariance matrix that depends on $\hat{\mu}$ and the counts in the marginals $\left\{ n_{C}:C\in \mathcal{C}\right\}$. Algorithm \ref{alg:boothbutler} outlines the method for sampling tables from $T_{\delta}$ introduced by Booth and Butler \shortcite{booth-butler-1999}.

\begin{algorithm}
\caption{Sample a table $n^{\prime}_{\delta}\in T_{\delta}$ (Booth and Butler, 1999)}
\label{alg:boothbutler}
\begin{algorithmic}[1]
\STATE Consider a table $n^{\prime}_{\delta}$ whose cells are currently unoccupied.
\STATE Partition the cells as bound $x_{\delta}^{B}$ and free $x_{\delta}^{F}$. 
\STATE Sample $\gamma_{1} \sim \normal\left((\hat{\mu}_{\delta})_{1},\hat{v}^{\delta}_{11}\right)$, the marginal distribution of $(x_{\delta})^{F}_{1}$ as derived from the joint distribution (\ref{eq:mvnfree}).
\STATE Set $(n^{\prime}_{\delta})^{F}_{1} = [\gamma_{1}]$. Here $[a]$ represents the nearest integer to $a$.
\FOR{$j=2,\ldots,m_{c}-m_{r}^{\prime}$}
\STATE Sample from the marginal distribution of $(x_{\delta})^{F}_{j}$ conditional on the current values of the preceding free cells as derived from the joint distribution (\ref{eq:mvnfree}):
$$\hspace{-1cm}\gamma_{j} \sim \normal\left( \E[(x_{\delta})^{F}_{j}\mid (x_{\delta})^{F}_{(1:(j-1))}=(n_{\delta}^{\prime})^{F}_{(1:(j-1))}],\V[(x_{\delta})^{F}_{j}\mid (x_{\delta})^{F}_{(1:(j-1))}=(n_{\delta}^{\prime})^{F}_{(1:(j-1))}]\right).$$
\STATE Set $(n_{\delta}^{\prime})^{F}_{j} = [\gamma_{j}]$.
\ENDFOR
\STATE Determine the values of the bound cells $(n^{\prime}_{\delta})^{B}$ using equation (\ref{eq:boundsfromfree}).
\IF {$(n^{\prime}_{\delta})^{B}$ does not contain negative entries}
\RETURN{ the table $n^{\prime}_{\delta}\in T_{\delta}$ determined by $(n^{\prime}_{\delta})^{F}$ and $(n^{\prime}_{\delta})^{B}$}
\ELSE
\STATE STOP. \COMMENT{The algorithm terminates without generating a table} 
\ENDIF
\end{algorithmic}
\end{algorithm}

It is worthwhile to compare how Algorithms \ref{alg:randomtablerref} and  \ref{alg:boothbutler} differ. A contingency table in $T_{\delta}$ is determined in Algorithm \ref{alg:randomtablerref} by sequentially calculating lower and upper bounds associated with the free cell whose value is sampled next, which entails solving $2(m_{c}-m_{r}^{\prime})$ optimization problems. In Algorithm \ref{alg:boothbutler} the calculation of bounds is replaced by simulations from multivariate normal distributions whose means and variances are obtained through fast matrix operations \cite{booth-butler-1999}. Since the determination of bounds comes at a higher computational cost, Algorithm \ref{alg:boothbutler} is much faster than Algorithm \ref{alg:randomtablerref}. Unfortunately, Algorithm \ref{alg:boothbutler} gives no guarantees that it will actually identify any table in $T_{\delta}$. A necessary condition for the successful generation of a table in $T_{\delta}$ is that the values sampled at lines 3 and 6 of Algorithm \ref{alg:boothbutler} are actually between their lower and upper bounds calculated at line 3 of Algorithm \ref{alg:randomtablerref}. To the best of the authors' knowledge, there does not exist any proof of this claim. In fact, this necessary condition is not mentioned in Booth and Butler \shortcite{booth-butler-1999} or in the subsequent work of Caffo and Booth \shortcite{caffo-booth-2001}. From a theoretical perspective, there is no justification why Algorithm \ref{alg:boothbutler} should successfully output a feasible table in $T_{\delta}$. Furthermore, there is no justification why Algorithm \ref{alg:boothbutler} should be able to sample any table in $T_{\delta}$ with strictly positive probability. Despite being faster, Algorithm \ref{alg:boothbutler} should not be preferred to Algorithm \ref{alg:randomtablerref} due to its lack of theoretical underpinning. In addition, Algorithm \ref{alg:randomtablerref} can be used to sample from reference sets of tables defined by bounds constraints in addition to linear constraints induced by fixed marginals, while Algorithm \ref{alg:boothbutler} cannot be used in such general situations because it relies on the calculations of MLEs associated with a log-linear model.\\
\indent Algorithm \ref{alg:boothbutler} is employed by Booth and Butler \shortcite{booth-butler-1999} to develop an importance sampling approach for producing Monte Carlo estimates of exact p-values. Caffo and Booth \shortcite{caffo-booth-2001} slightly modify Algorithm \ref{alg:boothbutler} by fixing a random number of free cells to develop a Markov chain algorithm for conditional inference. Both papers present successful applications of Algorithm \ref{alg:boothbutler} in generating feasible tables from a reference set $T$. However, such examples cannot substitute the need to provide rigorous proofs justifying the applicability of Algorithm \ref{alg:boothbutler}. Without such proofs one cannot know when to expect Algorithm \ref{alg:boothbutler} to succeed or to fail.\\
\indent For these reasons the existent literature does not seem to contain a reliable method for calculating exact p-values that works for arbitrary multi-way tables subject to linear and bounds constraints. In the next section we propose a new Markov chain algorithm that makes use of Algorithm \ref{alg:randomtablerref} to sample tables from a reference set. There is a significant advantage of using  Algorithm \ref{alg:randomtablerref} in the context of a Markov chain algorithm as opposed to an importance sampling procedure such as  SIS: the discrete distribution (\ref{eq:sampleneighbor}) becomes a proposal distribution for generating the candidate for the next state of the chain. The accuracy of the resulting exact p-values estimates is tied significantly less to how close the discrete distribution (\ref{eq:sampleneighbor}) is to the hypergeometric or uniform target distributions. Moreover, the instances in which Algorithm \ref{alg:randomtablerref} ends without successfully generating a table in the reference set are thrown out in an importance sampling method. On the other hand, a Markov chain procedure makes use of all the output from Algorithm \ref{alg:randomtablerref} even if no feasible table was identified.

\section{The Proposed Markov Chain Algorithm} \label{sec:markovchain}

\noindent We present a Markov chain algorithm that samples from a distribution $P_{*}(\cdot )$ on the reference set of tables $T$ whose key component is the dynamic Markov bases introduced in Section \ref{sec:dmb}. Algorithm \ref{alg:randomtablerref} generates feasible tables given an ordering of the cell indices $\mathcal{I}$ induced by a permutation $\delta\in \Delta_{m_{c}}$. The partitioning of the cells as bound and free as well as the sequence in which the values of the free cells are sampled are a function of the choice of $\delta$. The linear and bounds constraints that define $T$ and the sequence of free cells associated with permutations $\delta$ translate into various lower and upper bounds for the possible values of a particular cell. Empirically we observed that some tables in $T$ receive very high probabilities (\ref{eq:sampleneighbor}) of being sampled under some permutations in $\Delta_{m_{c}}$, but under other permutations the same probabilities could be very low. Characterizing the relationship between the discrete distribution (\ref{eq:sampleneighbor}) and a distribution on $T$ as a function of various cell orderings and distributions $f^{(L:U)}(\cdot)$ is a difficult problem that is currently open. The mixing time of a Markov chain that calls Algorithm \ref{alg:randomtablerref} to generate candidate tables could vary considerably if the permutation $\delta\in \Delta_{m_{c}}$ remains fixed across iterations. Since there are no theoretical results that would allow one to produce cell orderings that lead to smaller mixing times, we develop a Markov chain with state space $T\mytimes \Delta_{m_{c}}$ with stationary distribution
\begin{eqnarray} \label{eq:jointstationary}
 \Pr( n,\delta) & = & \Pr(n\mid \delta)\Pr(\delta).
\end{eqnarray}
Conditional on $\delta \in  \Delta_{m_{c}}\setminus \{\delta_{0}\}$, a table $n\in T=T_{\delta_{0}}$ is transformed in a table $n_{\delta}\in T_{\delta}$ with the same cell counts but a different ordering of its cells. This implies $\Pr(n\mid \delta) = \Pr(n_{\delta})$. Sampling from the joint distribution (\ref{eq:jointstationary}) is relevant in this context only if $P_{*}(\cdot )$ coincides with the marginal distribution $\Pr(n) = \sum_{\delta\in \Delta_{m_{c}}}\Pr(n_{\delta})\Pr(\delta)$. This condition is satisfied for $\Pr(n_{\delta})=P_{*}(n_{\delta})$ if $P_{*}(\cdot )$ is invariant to cell orderings, i.e. $P_{*}(n_{\delta}) = P_{*}(n)$ for all $\delta \in \Delta_{m_{c}}$. The uniform and hypergeometric distributions $P_{U}(\cdot )$ and $P_{H}(\cdot )$ from equation (\ref{eq:targetdistrib}) are indeed order invariant. Since there is no reason to favor a cell ordering over another, we assume a uniform distribution $\Uni_{\Delta_{m_{c}}}(\cdot)$ on the set of possible permutations of $\mathcal{I}$. Therefore the stationary distribution (\ref{eq:jointstationary}) is 
\begin{eqnarray} \label{eq:jointstationarygood}
 \Pr( n,\delta) & = & P_{*}(n_{\delta})\Uni_{\Delta_{m_{c}}}(\delta).
\end{eqnarray}
We note that the marginal distribution of (\ref{eq:jointstationarygood}) associated with $\delta$ is again uniform.\\
\indent We start the chain by sampling a permutation $\delta^{(0)} \sim \Uni_{\Delta_{m_{c}}}(\cdot)$ and using Algorithm \ref{alg:randomtablerref} with cell ordering $\delta^{(0)}$ to sample a table $n^{(0)} \in T$. Given a current state $(n^{(t)},\delta^{(t)})$, we sample a new permutation $\delta^{(t+1)} \sim \Uni_{\Delta_{m_{c}}}(\cdot)$. We also sample a candidate table $n^{*}_{\delta^{(t+1)}}\in T_{\delta^{(t+1)}}$ from a proposal distribution $q_{\delta^{(t+1)}}(n^{(t)}_{\delta^{(t+1)}},\cdot)$. If we did not obtain a feasible table (i.e., $n^{*}\notin T$), the next state of the chain is $(n^{(t)},\delta^{(t+1)})$. If $n^{*}\in T$, the next state $(n^{(t+1)},\delta^{(t+1)})$ is $(n^{*},\delta^{(t+1)})$ with the Metropolis-Hastings probability
\begin{eqnarray}\label{eq:acceptanceprob}
 \min\left\{ 1, \frac{P_{*}(n^{*}_{\delta^{(t+1)}})q_{\delta^{(t+1)}}(n^{*}_{\delta^{(t+1)}},n^{(t)}_{\delta^{(t+1)}})}{P_{*}(n^{(t)}_{\delta^{(t+1)}})q_{\delta^{(t+1)}}(n^{(t)}_{\delta^{(t+1)}},n^{*}_{\delta^{(t+1)}})}\right\}.
\end{eqnarray}
Otherwise we set $n^{(t+1)}=n^{(t)}$. A sufficient condition for the irreducibility of this Markov chain is the positivity of the instrumental distribution, i.e.
\begin{eqnarray} \label{eq:positivity}
	q_{\delta}(n^{\prime}_{\delta},n^{\prime\prime}_{\delta})>0, \mbox{ for every }(n^{\prime}_{\delta},n^{\prime\prime}_{\delta})\in T_{\delta}\times T_{\delta} \mbox{ and } \delta\in \Delta_{m_{c}}.
\end{eqnarray}
It is possible to employ Algorithm \ref{alg:randomtablerref} to generate candidate tables $n^{*}$. In this case the Markov chain stays at its current state if Algorithm \ref{alg:randomtablerref} does not generate a feasible table in $T$. If a feasible candidate table $n^{*}$ is identified, the acceptance probability (\ref{eq:acceptanceprob}) is calculated based on the proposal distribution $q_{\delta^{(t+1)}}(n^{(t)}_{\delta^{(t+1)}},n^{*}_{\delta^{(t+1)}}) = \pi_{\delta^{(t+1)},f^{(L:U)}(\cdot)}(n^{*}_{\delta^{(t+1)}})$ \--- see equation  (\ref{eq:sampleneighbor}). Since Algorithm \ref{alg:randomtablerref} can return any table in $T$, the positivity condition (\ref{eq:positivity}) is satisfied. Unfortunately the probability of proposing a candidate table at iteration $t$ from the reference set is independent of $n^{(t)}$. This leads to an erratic behavior of the Markov chain with very small acceptance rates for new candidate tables.\\
\indent A better option is to sample candidate tables $n^{*}$ that have a number $M$ of cell counts in common with $n^{(t)}$. The maximum value for $M$ is $(m_{c}-m^{\prime}_{r}-1)$. Recall that $(m_{c}-m^{\prime}_{r})$ is the number of free cells associated with $T$. For a permutation $\delta\in \Delta_{m_{c}}$, we construct a proposal distribution $q_{\delta}(n^{(t)}_{\delta},\cdot)$ as follows. We partition $T_{\delta}\setminus \{n_{\delta}^{(t)}\}$ in subsets of tables that have the counts of the first $M$ free cells equal with the counts of the first $M$ free cells of $n^{(t)}_{\delta}$ but the count of the $(M+1)$-th free cell different than $(n^{(t)}_{\delta})^{F}_{M+1}$:
\begin{eqnarray}\label{eq:tdeltadecomp}
 T_{\delta}\setminus\{n_{\delta}^{(t)}\} & = &\mycup\limits_{M=0}^{m_{c}-m^{\prime}_{r}-1} \mbox{nbd}_{\delta,M}(n^{(t)}_{\delta}),
 \end{eqnarray}
\noindent where 
\begin{eqnarray*}
\mbox{nbd}_{\delta,M}(n^{(t)}_{\delta}) & = & T^{\prime}_{\delta,M}(n^{(t)}_{\delta})\setminus T^{\prime}_{\delta,M+1}(n^{(t)}_{\delta}), \mbox{ and }\\
 T^{\prime}_{\delta,M}(n^{(t)}_{\delta}) & = & \left\{ n^{\prime}_{\delta}\in T_{\delta}:(n^{\prime}_{\delta})^{F}_{j}=(n^{(t)}_{\delta})^{F}_{j}, \mbox{ for }j=1,\ldots,M\right\}.
\end{eqnarray*}
 \noindent We remark that $T^{\prime}_{\delta,m_{c}-m^{\prime}_{r}}(n^{(t)}_{\delta})=\{n^{(t)}_{\delta}\}$. We refer to the tables in $\mbox{nbd}_{\delta,M}(n^{(t)}_{\delta})$ as the neighbors of $n^{(t)}_{\delta}$ of order $M$. Any table in $T_{\delta}\setminus \{n_{\delta}^{(t)}\}$  is the neighbor of $n^{(t)}_{\delta}$ of a particular order. For certain values of $M\in (0:(m_{c}-m^{\prime}_{r}-1))$, there might not exist a neighbor of $n^{(t)}_{\delta}$ of order $M$.\\
 \indent We modify Algorithm \ref{alg:randomtablerref} into Algorithm \ref{alg:mcmctable} such that the feasible tables it generates are neighbors of order $M$ of table $n^{(t)}_{\delta}$. Line 2 of Algorithm \ref{alg:mcmctable} guarantees that, if a feasible table is returned, then this table belongs to  $T^{\prime}_{\delta,M}(n^{(t)}_{\delta})$. By eliminating $(n^{(t)}_{\delta})^{F}_{M+1}$ from the possible values of the free cell $(n^{\prime}_{\delta})^{F}_{M+1}$ in line 3, we guarantee that Algorithm \ref{alg:mcmctable} does not return a table that belongs to $T^{\prime}_{\delta,M+1}(n^{(t)}_{\delta})$. Algorithm \ref{alg:mcmctable} samples a table $n^{\prime}_{\delta}\in  \mbox{nbd}_{\delta,M}(n^{(t)}_{\delta})$ with strictly positive probability:
\begin{eqnarray} \label{eq:probM}
 \pi^{F}_{\delta,f^{(L:U)}(\cdot),M} (n^{\prime}_{\delta}\mid n^{(t)}_{\delta}) \propto f^{\left(L^{F}_{\delta,n^{\prime},M}:U^{F}_{\delta,n^{\prime},M}\right)\setminus \{ (n^{(t)}_{\delta})^{F}_{M+1}\}}((n^{\prime}_{\delta})^{F}_{M+1})  \prod\limits_{j=M+2}^{m_{c}-m^{\prime}_{r}} f^{(L^{F}_{\delta,n^{\prime},j}:U^{F}_{\delta,n^{\prime},j})}((n^{\prime}_{\delta})^{F}_{j}).
\end{eqnarray}
 
\algsetup{linenosize=\small,linenodelimiter=.}

\begin{algorithm}
\caption{Sample a table $n^{\prime}_{\delta}\in  \mbox{nbd}_{\delta,M}(n^{(t)}_{\delta})$}
\label{alg:mcmctable}
\begin{algorithmic}[1]
\STATE Consider a table $n^{\prime}_{\delta}$ whose cells are currently unoccupied.
\STATE Set the counts of the first $M$ free cells of $n^{\prime}_{\delta}$ to the corresponding counts of $n^{(t)}_{\delta}$, i.e.
$$
 (n^{\prime}_{\delta})^{F}_{j} = (n^{(t)}_{\delta})^{F}_{j}, \mbox{ for } j=1,\ldots,M.
$$
\STATE Sample the values of the remaining free cells $(n^{\prime}_{\delta})^{F}_{j}$, $j=M+1,\ldots,m_{c}-m_{r}^{\prime}$ using line 3 of Algorithm \ref{alg:randomtablerref}. When sampling the value of the $(M+1)$-th free cell, eliminate $(n^{(t)}_{\delta})^{F}_{M+1}$ from the set of possible values of this cell.
\STATE Attempt to determine a full table $n^{\prime}_{\delta}$ as described in lines 5-12 of Algorithm \ref{alg:randomtablerref}.
\end{algorithmic}
\end{algorithm}

\indent We consider the set of all the local moves associated with $n^{(t)}_{\delta}$ in $T_{\delta}$:
\begin{eqnarray} \label{eq:mdelta}
 \mathcal{M}_{\delta}(n^{(t)}_{\delta}) & = &\left\{ n^{\prime}_{\delta}-n^{(t)}_{\delta}:n^{\prime}_{\delta}\in T_{\delta}\setminus\{n_{\delta}^{(t)}\}\right\}.
\end{eqnarray}
Their union $\mycup_{n_{\delta}^{(t)}\in T_{\delta}}\mathcal{M}_{\delta}(n^{(t)}_{\delta})$ is a Markov basis for $T_{\delta}$.  The decomposition (\ref{eq:tdeltadecomp}) of $T_{\delta}\setminus\{n_{\delta}^{(t)}\}$ as sets of neighbor tables of $n_{\delta}^{(t)}$ of various orders translates into a corresponding decomposition of the set of local moves associated with $n^{(t)}_{\delta}$ in $T_{\delta}$:
\begin{eqnarray}\label{eq:localmovesdec}
 \mathcal{M}_{\delta}(n^{(t)}_{\delta}) = \mycup\limits_{M=0}^{m_{c}-m^{\prime}_{r}-1} \mathcal{M}_{\delta,M}(n^{(t)}_{\delta}),
\end{eqnarray}
\noindent where $\mathcal{M}_{\delta,M}(n^{(t)}_{\delta})= \left\{ n^{\prime}_{\delta}-n^{(t)}_{\delta}:n^{\prime}_{\delta}\in \mbox{nbd}_{\delta,M}(n^{(t)}_{\delta})\right\}$. We dynamically generate local moves in $\mathcal{M}_{\delta}(n^{(t)}_{\delta})$ as follows. We consider a discrete distribution $g_{T}(\cdot)$ that gives a strictly positive probability to each integer in $(0:(m_{c}-m^{\prime}_{r}-1))$. We draw $M\sim g_{T}(\cdot)$ then employ Algorithm \ref{alg:mcmctable} to sample a table $n^{*}_{\delta}\in \mbox{nbd}_{\delta}(n^{(t)}_{\delta})$. This gives us a local move $n^{*}_{\delta}-n^{(t)}_{\delta}\in \mathcal{M}_{\delta,M}(n^{(t)}_{\delta})$ without having to determine the entire set $\mathcal{M}_{\delta,M}(n^{(t)}_{\delta})$. We use this procedure to sample candidate tables for the Markov chain algorithm with stationary distribution (\ref{eq:jointstationarygood}). The corresponding instrumental distribution is given by
\begin{eqnarray} \label{eq:proposalmixture}
 q_{\delta}(n^{(t)}_{\delta},n^{*}_{\delta}) & = & \sum_{M=0}^{m_{c}-m^{\prime}_{r}-1} g_{T}(M) \pi^{F}_{\delta,f^{(L:U)}(\cdot),M} (n^{*}_{\delta}\mid n^{(t)}_{\delta}) I_{\left\{n^{*}_{\delta}\in \mbox{nbd}_{\delta,M}(n^{(t)}_{\delta})\right\}}.
\end{eqnarray}
\noindent We remark that $n^{*}_{\delta}\in \mbox{nbd}_{\delta,M}(n^{(t)}_{\delta})$ implies $n^{(t)}_{\delta}\in \mbox{nbd}_{\delta,M}(n^{*}_{\delta})$ and $n^{(t)}_{\delta}\notin \mbox{nbd}_{\delta,M^{\prime}}(n^{*}_{\delta})$ for $M^{\prime}\ne M$. Therefore, in order to calculate the Metropolis-Hastings acceptance ratio (\ref{eq:acceptanceprob}), we need to evaluate only one component of the mixture distribution (\ref{eq:proposalmixture}). For $n^{*}_{\delta}\in \mbox{nbd}_{\delta,M}(n^{(t)}_{\delta})$, we have
\begin{eqnarray*}
 \frac{q_{\delta}(n^{*}_{\delta},n^{(t)}_{\delta})}{q_{\delta}(n^{(t)}_{\delta},n^{*}_{\delta})} & = & \frac{\pi^{F}_{\delta,f^{(L:U)}(\cdot),M} (n^{(t)}_{\delta}\mid n^{*}_{\delta})}{\pi^{F}_{\delta,f^{(L:U)}(\cdot),M} (n^{*}_{\delta}\mid n^{(t)}_{\delta})}.
\end{eqnarray*}
\noindent This makes the computing effort needed to run the resulting Markov chain quite manageable. The chain is irreducible because the positivity condition (\ref{eq:positivity}) is satisfied as a result of $\mycup_{n_{\delta}^{(t)}\in T_{\delta}}\mathcal{M}_{\delta}(n^{(t)}_{\delta})$ being a Markov basis for $T_{\delta}$.\\
\indent The choice of the discrete distribution $g_{T}(\cdot)$ is crucial for a good performance of the chain. The values of $M$ sampled from $g_{T}(\cdot)$ need to maintain a balance between making large jumps in $T$ (hence being more likely to reject the move) and making small jumps in $T$ (hence being less likely to reject the move, but spending many iterations around similar tables). Specifying a reasonable distribution $g_{T}(\cdot)$ could be a daunting task since it needs to be tailored specifically for $T$. In the next section we give a coherent procedure for finding $g_{T}(\cdot)$ based on a flexible algorithm for exploring an arbitrary target set of tables.

\section{The Algorithm for Finding $g_{T}(\cdot)$} \label{sec:finding}

\noindent We present a method for producing a discrete distribution $g_{T}(\cdot)$ required in the specification of the proposal distribution (\ref{eq:proposalmixture}). Our approach is based on a repeated approximation of the number of free cells whose counts need to be fixed before all the other counts are uniquely determined. An upper bound for this number is the total number of free cells $(m_{c}-m_{r}^{\prime})$ \--- see Section \ref{sec:dmb}. However, due to the particular configurations of small and large counts of the tables in $T$ and to the presence of the bounds constraints (\ref{eq:boundsconstraintsT}), this number can actually be anywhere between $1$ and $(m_{c}-m_{r}^{\prime})$.\\
\indent We assume that $T$ contains at least two tables. We consider a permutation $\delta\in \Delta_{m_{c}}$ and a table $n^{*}\in T$. We let $\mathcal{F}_{\delta}\subset \mathcal{I}$  be the indices of the $(m_{c}-m_{r}^{\prime})$ free cells associated with $T$ and $\delta$ \--- see Section \ref{sec:dmb}. We take $\delta^{\prime}\in \Delta_{m_{c}}$ such that $\left\{i^{\delta^{\prime}(1)},i^{\delta^{\prime}(2)},\ldots,i^{\delta^{\prime}(m_{c}-m_{r}^{\prime})}\right\}=\mathcal{F}_{\delta}$. Recall from equation (\ref{eq:tdelta}) that $T_{\delta^{\prime},s}(n^{*}_{\delta^{\prime}})$ represents the set of tables in $T$ that have the same counts in cells $\{i^{\delta^{\prime}(1)},\ldots,i^{\delta^{\prime}(s)}\}$ as table $n^{*}$. There exists a unique index $s(\delta^{\prime},n^{*})\in (0:(m_{c}-m_{r}^{\prime}-1))$ such that
\begin{itemize}
\item[] (C1) $T_{\delta^{\prime},s(\delta^{\prime},n^{*})}(n^{*}_{\delta^{\prime}})$ contains at least one table in $T$ that is different than $n^{*}$; 
\item[] (C2) $T_{\delta^{\prime},s(\delta^{\prime},n^{*})+1}(n^{*}_{\delta^{\prime}})=\{n^{*}_{\delta^{\prime}}\}$.
\end{itemize}
If $T_{\delta^{\prime},s}(n^{*}_{\delta^{\prime}})$ contains a table different than $n^{*}_{\delta^{\prime}}$, there must exist $j\in ((s+1):(m_{c}-m_{r}^{\prime}))$ such that the lower and upper bounds of the corresponding cell are different:
\begin{eqnarray} \label{eq:boundsdifferent}
  \left\lceil L^{R}_{\delta^{\prime},n^{*},s}(i^{\delta^{\prime}(j)})\right\rceil < \left\lfloor U^{R}_{\delta^{\prime},n^{*},s}(i^{\delta^{\prime}(j)})\right\rfloor.
\end{eqnarray}
This condition is necessary, but it is not sufficient. That is, equation (\ref{eq:boundsdifferent}) might hold while still  $T_{\delta^{\prime},s}(n^{*}_{\delta^{\prime}})=\{n^{*}_{\delta^{\prime}}\}$. As such, the computation of real bounds cannot substitute actually checking that $T_{\delta^{\prime},s}(n^{*}_{\delta^{\prime}})\setminus \{n^{*}_{\delta^{\prime}}\}\ne \emptyset$, but such a check is computationally expensive. We reduce this computing effort by first determining an upper bound for $s(\delta^{\prime},n^{*})$.\\
\indent Algorithm \ref{alg:binarysearch} performs a binary search to determine the maximum index $s^{b}(\delta^{\prime},n^{*})< (m_{c}-m_{r}^{\prime})$ such that
\begin{eqnarray} \label{eq:boundsequal}
 \left\lceil L^{R}_{\delta^{\prime},n^{*},s^{b}(\delta^{\prime},n^{*})+1}(i^{\delta(j)})\right\rceil = n^{*}(i^{\delta(j)})= \left\lfloor U^{R}_{\delta^{\prime},n^{*},s^{b}(\delta^{\prime},n^{*})+1}(i^{\delta(j)})\right\rfloor,
\end{eqnarray}
\noindent for $j\in ((s^{b}(\delta^{\prime},n^{*})+2):(m_{c}-m_{r}^{\prime}))$, but $\left\lceil L^{R}_{\delta^{\prime},n^{*},s^{b}(\delta^{\prime},n^{*})}(i^{\delta(j)})\right\rceil < \left\lfloor U^{R}_{\delta^{\prime},n^{*},s^{b}(\delta^{\prime},n^{*})}(i^{\delta(j)})\right\rfloor$ for at least one index $j\in ((s^{b}(\delta^{\prime},n^{*})+1):(m_{c}-m_{r}^{\prime}))$. Condition (\ref{eq:boundsdifferent}) is always satisfied for $s=0$ as long as $T$ contains at least two tables. Moreover, condition (\ref{eq:boundsdifferent}) is never satisfied for $s=m_{c}-m_{r}^{\prime}$ since $T_{\delta^{\prime},m_{c}-m_{r}^{\prime}}(n^{*}_{\delta^{\prime}})=\{ n^{*}_{\delta^{\prime}}\}$. At the completion of Algorithm \ref{alg:binarysearch}, the index $s^{b}(\delta^{\prime},n^{*})$ is returned and we still need to determine $s(\delta^{\prime},n^{*})$. Since $T_{\delta^{\prime},s^{b}(\delta^{\prime},n^{*})+1}(n^{*}_{\delta^{\prime}})=\{n^{*}_{\delta^{\prime}}\}$, condition (C2) is satisfied and hence $s(\delta^{\prime},n^{*})\le s^{b}(\delta^{\prime},n^{*})$. Algorithm \ref{alg:sdelta} starts with $s^{b}(\delta^{\prime},n^{*})$ as the initial guess for the value of $s(\delta^{\prime},n^{*})$ and sequentially decreases this guess until condition (C1) is also satisfied.\\

\algsetup{linenosize=\small,linenodelimiter=.}

\begin{algorithm}
\caption{Determination of $s^{b}(\delta^{\prime},n^{*})$}
\label{alg:binarysearch}
\begin{algorithmic}[1]
\STATE Set $s_{1} \leftarrow 0$ and $s_{2} \leftarrow (m_{c}-m_{r}^{\prime})$.
\WHILE{$s_{1}+1\ne s_{2}$}
\STATE Set $s \leftarrow \lfloor(s_{1}+s_{2})/2\rfloor$.
\FOR{$j=s+1,\ldots,m_{c}-m_{r}^{\prime}$}
\STATE Calculate the real lower and upper bounds $L^{R}_{\delta^{\prime},n^{*},s}(i^{\delta(j)})$ and $U^{R}_{\delta^{\prime},n^{*},s}(i^{\delta(j)})$.
\ENDFOR
\IF{$\left\lceil L^{R}_{\delta^{\prime},n^{*},s}(i^{\delta(j)})\right\rceil = \left\lfloor U^{R}_{\delta^{\prime},n^{*},s}(i^{\delta(j)}) \right\rfloor$ for all  $j=s+1,\ldots,m_{c}-m_{r}^{\prime}$}
\STATE Set $s_{2}\leftarrow s$.
\ELSE
\STATE Set $s_{1}\leftarrow s$.
\ENDIF
\ENDWHILE
\STATE Return $s^{b}(\delta^{\prime},n^{*})\leftarrow s_{1}$.
\end{algorithmic}
\end{algorithm}
\begin{algorithm}
\caption{Determination of a lower bound of $s(\delta^{\prime},n^{*})$}
\label{alg:sdelta}
\begin{algorithmic}[1]
 \FOR{$s=s^{b}(\delta^{\prime},n^{*}),s^{b}(\delta^{\prime},n^{*})-1,\ldots,0$}
\STATE Set free cells $\{ i^{\delta^{\prime}(1)},\ldots,i^{\delta^{\prime}(s)}\}$ to the corresponding values from $n^{*}$.
\STATE Sample the values of the remaining components of the vector of free cells $x_{\delta}^{F}$ using lines 5-14 of Algorithm \ref{alg:randomtable}.
\STATE Attempt to determine a full table $n^{\prime}\in T$ as described in lines 5-12 of Algorithm \ref{alg:randomtablerref}.
\IF{a table $n^{\prime}\in T$ was determined}
	\IF{$n^{\prime}\ne n^{*}$}
 		\RETURN{$s$}
 	\ENDIF
\ENDIF
 \ENDFOR
\end{algorithmic}
\end{algorithm}
\indent Algorithm \ref{alg:sdelta} returns a value of $s$ that is less or equal than $s(\delta,n^{\prime})$. However, for our purposes, this lower bound is sufficient. Algorithm \ref{alg:gdot} estimates the discrete distribution $g_{T}(\cdot)$ by repeatedly calling Algorithms \ref{alg:randomtablerref}, \ref{alg:binarysearch} and \ref{alg:sdelta} for a large number of iterations $i_{max}$. The value of $g_{T}(j)$, $j\in (0:(m_{c}-m_{r}^{\prime}-1))$, is proportional with the number of iterations in which keeping $j$ counts of free cells fixed resulted in the successful sampling of a feasible table different than some other randomly generated feasible table. The initialization from line 2 of Algorithm \ref{alg:gdot} assures that the distribution $g_{T}(\cdot)$ returned by the procedure satisfies $g_{T}(j) > 0$ for any $j\in (0:(m_{c}-m_{r}^{\prime}-1))$ which is a condition required to generate all the local moves from equation (\ref{eq:mdelta}). Algorithm \ref{alg:gdot} effectively explores the set of tables $T$ and identifies a distribution $g_{T}(\cdot)$ based on this exploration. We remark that we have not made any assumptions about a parametric form for $g_{T}(\cdot)$. The structure of $T$ dictates the probabilities that define $g_{T}(\cdot)$ which leads to a very flexible choice of $g_{T}(\cdot)$ which is adapted to the structure of $T$.

\begin{algorithm}
\caption{Determination of $g_{T}(\cdot)$}
\label{alg:gdot}
\begin{algorithmic}[1]
 \STATE Consider a vector $G$ with indices $(1:(m_{c}-m_{r}^{\prime}))$.
 \STATE Set $G(j) \leftarrow 1$ for $j\in (1:(m_{c}-m_{r}^{\prime}))$.
 \FOR{$i=1,2,\ldots,i_{max}$}
 \STATE Call Algorithm \ref{alg:randomtablerref} until it generates a random table $n^{*}\in T$.
 \STATE Generate a random permutation $\delta\in \Delta_{m_{c}}$ and find the RREF of the linear system (\ref{eq:linsysdelta}).
 \STATE Call Algorithm \ref{alg:binarysearch} to determine $s^{b}(\delta^{\prime},n^{*})$.
 \STATE Call Algorithm \ref{alg:sdelta} to determine $s\le s(\delta^{\prime},n^{*})$.
 \STATE Set $G(s) \leftarrow G(s)+1$.
 \ENDFOR
 \STATE Set $g_{T}(j) = G(j+1)/i_{max}$ for $j\in (0:(m_{c}-m_{r}^{\prime}-1))$.
 \RETURN{$g_{T}(\cdot)$}
\end{algorithmic}
\end{algorithm}

\section{Examples} \label{sec:examples}

\noindent We illustrate the use of the Markov chain algorithm with dynamic Markov bases in two examples. The first example involves a three-way table with structural zeros, while the second example involves a sparse eight-way table. Both examples have been chosen to show the effectiveness of the Markov chain algorithm described in Sections \ref{sec:markovchain} and \ref{sec:finding} with respect to competing approaches proposed in the literature. For both examples, we have been unable to generate a Markov basis using the computational algebraic techniques of Diaconis and Sturmfels \shortcite{diaconis-sturmfels-1998}, which renders their sampling approach inapplicable. The sequential importance sampling (SIS) algorithm of Chen et al. \shortcite{chen-annals2006} is applicable, but fails to provide any meaningful results by giving estimates equal to $1$ for all the p-values we calculate. The Markov chain algorithm of Caffo and Booth \shortcite{caffo-booth-2001} (CB, henceforth) as implemented in the R package \texttt{exactLoglinTest} \cite{caffo-2006} is not applicable for tables with structural zeros, hence it does not produce any estimates for our first example.\\
\indent We run $100$ independent Markov chains of length 2500000 with a burn-in time of 25000 iterations. The chains were run with the dynamic Markov bases approach and the CB algorithm. The SIS algorithm was run until it generated an equal number of sampled tables. We sampled from the hypergeometric distribution $P_{H}(\cdot)$ and calculated estimates for the exact p-values associated with the $X^{2}$ and $G^{2}$ statistics for the all two-way interaction model. We run $100$ replicates of Algorithm \ref{alg:gdot} for $100000$ iterations to find  the distribution $g_{T}(\cdot)$ that defines the Metropolis-Hastings proposal distribution (\ref{eq:proposalmixture}).\\
\indent We estimate the Monte Carlo error using the non-overlapping batch means method of Geyer \shortcite{geyer-1992}. Each of the $100$ independent chains was divided in $10$ batches of size $250000$. The standard error of an exact p-value estimate is the sample standard error of the p-value estimates corresponding with the $1000$ resulting batches. The Monte Carlo errors are given after the ``$\pm$'' sign following the Monte Carlo estimate of the exact p-value. We report the computing time necessary to generate one batch of $250000$ iterations throughout. We use OpenMPI (\texttt{http://www.open-mpi.org/}) to obtain batches by running independent processes on several processors.\\
\indent We performed our computations on a Mac Pro computer with 2 x 2.26 GHz quad-core Intel Xeon processors with 16 GB of memory. We report the mean elapsed computing time in seconds with standard errors calculated across the replicates. We wrote our own C++ implementation of the SIS algorithm by following the description from Chen et al. \shortcite{chen-annals2006}. The tables have been sampled in SIS using Algorithm \ref{alg:randomtable} with cell values generated from the hypergeometric distribution $f_{h}(\cdot)$. We implemented the algorithms described in Sections \ref{sec:dmb}, \ref{sec:markovchain} and \ref{sec:finding}  in C++. The linear programming problems (\ref{eq:linprogbounds}) have been solved with IBM ILOG CPLEX Optimizer (\texttt{http://www.ibm.com})  routines. All the code and the datasets needed to replicate the numerical results from this section are available as supplemental materials.

\subsection{NBER data}

\begin{figure}
 \centering
 \makebox{\includegraphics[width=6.5in,angle=0]{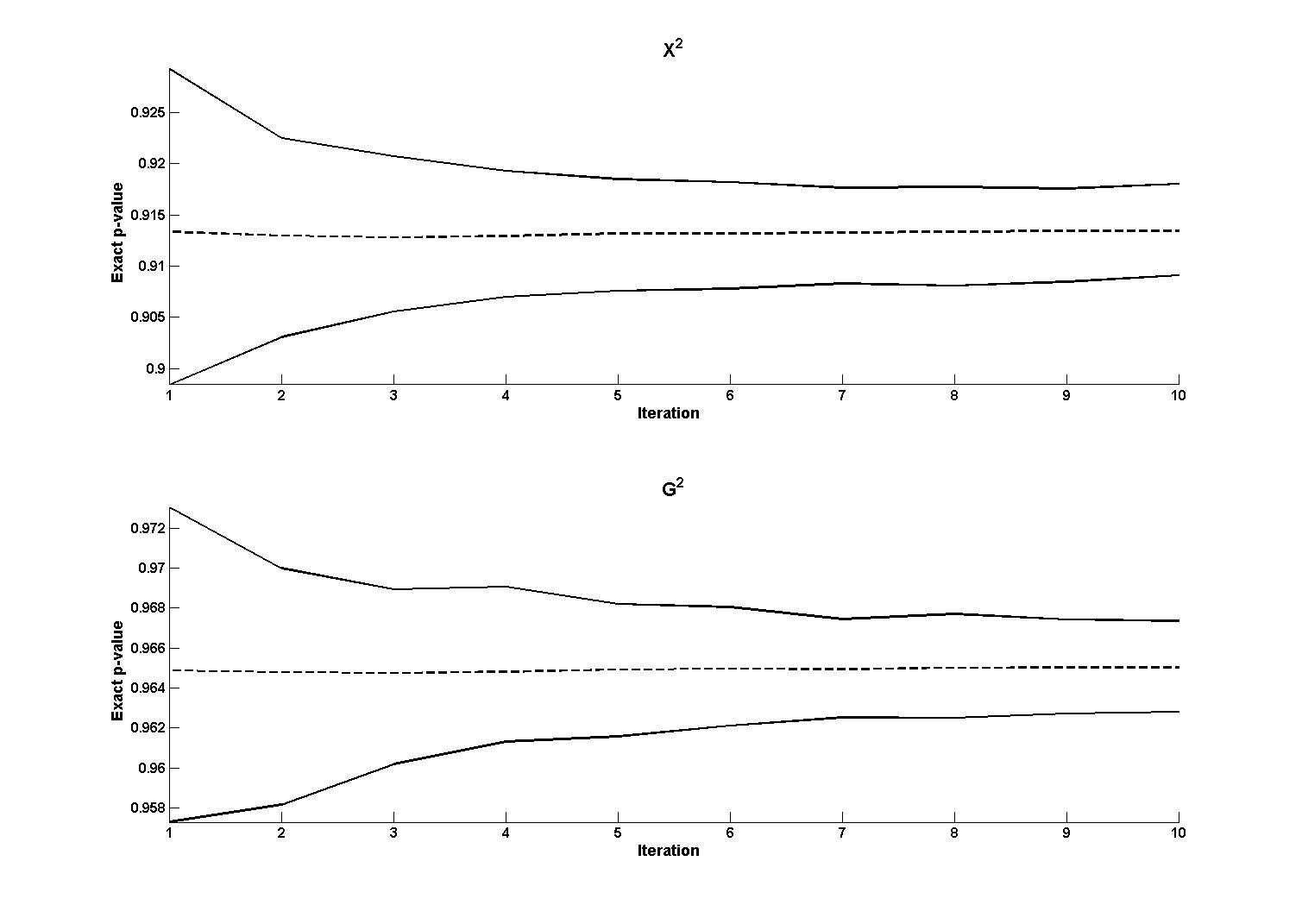}}
  \caption{\label{fig:nberconvergence} Convergence of $100$ independent Markov chains based on the dynamic Markov basis for the NBER data and the all two-way interaction model. The upper panel shows the convergence to the estimate $0.9134$ of the $X^{2}$ exact p-value, while the lower panel shows the convergence to the estimate $0.9650$ of the $G^{2}$ exact p-value. The $x$-axis gives the number of iterations in increments of $250000$. For each chain we calculated p-value estimates based on $250000i$ sampled tables with $i=1,2,\ldots,10$. The dotted line represents the mean of these incremental estimates, while the solid lines represent their $2.5\%$ and $97.5\%$ quantiles.}
\end{figure}

\begin{figure}
 \centering
 \makebox{\includegraphics[width=5in,angle=0]{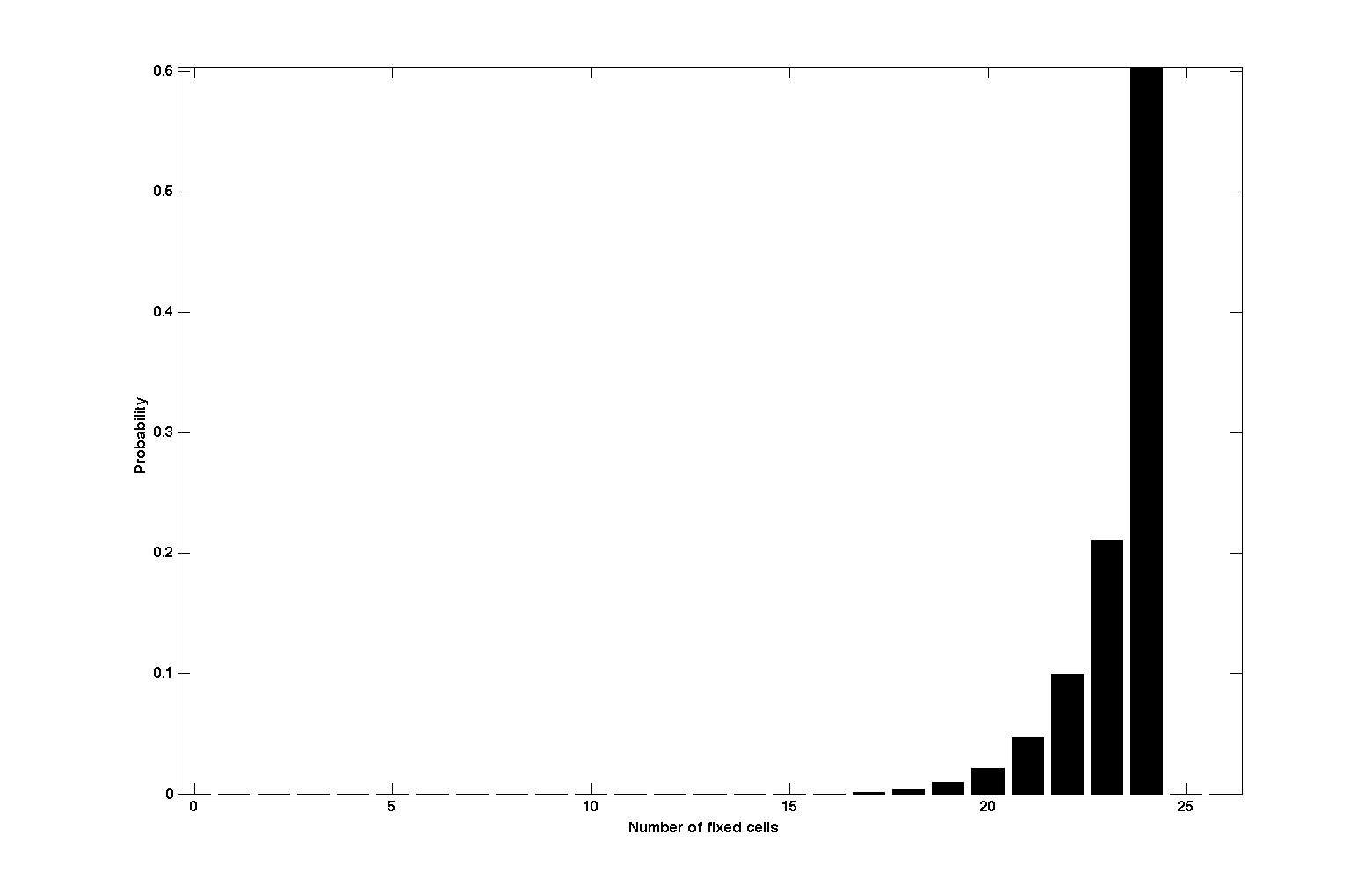}}
  \caption{\label{fig:nberdiagnostic} The discrete distribution $g_{T}(\cdot)$ for the NBER data and the all two-way interaction model as determined by Algorithm \ref{alg:gdot}.}
\end{figure}

\noindent Table \ref{table:nber} is a $4\times 5\times 4$ cross-classification of $4345$ individuals by occupational groups (O1 \--- ``self-employed, business'', O2 \--- ``self-employed, professional'', O3 \--- ``teacher'', O4 \--- ``salary-employed''), aptitude levels (A) and educational levels (E). It was collected in a 1969 survey of the National Bureau of Economic Research (NBER) \--- see Table 3-6 page 45 from Fienberg \shortcite{fienberg-book-2007}. The horizontal lines denote structural zeros. The ten structural zeros under O3 and E1, E2 are associated with teachers being required to have higher education levels. The other two structural zeros under O2 can be motivated in a similar manner.\\
\indent The number of degrees of freedom for the all two-way interaction model is calculated by subtracting the number of structural zeros from $36$ \--- the number of degrees of freedom corresponding with a $4\times 5\times 4$ table without structural zeros. Bishop et al. \shortcite{bishop75} argue that the number of degrees of freedom must be increased by the number of structural zeros that are present in marginal tables that are among the minimal sufficient statistics of the log-linear model considered. In this case there are two such counts present in the aptitude by educational levels marginal. The resulting number of degrees of freedom is $36-12+2=26$. The observed value of the likelihood-ratio test statistic is $G^2=15.91$ which leads to an asymptotic p-value for the all two-way interactions model of $0.938$. The observed value of the $X^2$ test statistic is $17.1$ which leads to an asymptotic p-value of $0.906$. \\
\indent The Markov chains with dynamic Markov bases lead to an estimate of the $G^{2}$ exact p-value of $0.9650\pm 0.0037$ and to an estimate of the $X^{2}$ exact p-value of $0.9134\pm 0.0068$. Figure \ref{fig:nberconvergence} shows the convergence of the Markov chain algorithm from Section \ref{sec:markovchain} across the $100$ chains.  We remark that the large sample size of the NBER data leads to a good agreement between the asymptotic and the exact p-values. The computing time for one batch of $250000$ tables for our Markov chain algorithm is $1882.58\pm 1.33$ seconds. Figure \ref{fig:nberdiagnostic} shows the estimated distribution function $g_{T}(\cdot)$ obtained from $10$ million iterations of Algorithm \ref{alg:gdot}. The number of free cells is $26$ hence its domain is $\{0,1,\ldots,25\}$. We see that the mode of $g_{T}(\cdot)$ is $g_{T}(24)=0.604$ which represents the estimated probability of obtaining a feasible table different than the current table after fixing the values of $24$ free cells. The running time of Algorithm \ref{alg:gdot} is $1266\pm 0.93
$ seconds per $100000$ iterations. 

\begin{table}[h]
\begin{center}
\caption{\label{table:nber} NBER data. The grand total of this table is $4345$.}
\begin{tabular}{ccccccccccccc}\hline
    &    & E1   & E2    & E3   & E4   & &    &    & E1   & E2    & E3  & E4\\ \hline
 O1 & A1 & $42$ & $55$  & $22$ & $3$  & & O3 & A1 & \--- & \--- & $1$ & $19$\\
    & A2 & $72$ & $82$  & $60$ & $12$ & &    & A2 & \--- & \--- & $3$ & $60$\\
    & A3 & $90$ & $106$ & $85$ & $25$ & &    & A3 & \--- & \--- & $5$ & $86$\\
    & A4 & $27$ & $48$  & $47$ & $8$  & &    & A4 & \--- & \--- & $2$ & $36$\\ 
    & A5 & $8$  & $18$  & $19$ & $5$  & &    & A5 & \--- & \--- & $1$ & $14$\\ \hline
 O2 & A1 & $1$  & $2$   & $8$  & $19$ & & O4 & A1 & $172$ & $151$ & $107$ & $42$\\
    & A2 & $1$  & $2$   & $15$ & $33$ & &    & A2 & $208$ & $198$ & $206$ & $92$\\
    & A3 & $2$  & $5$   & $25$ & $83$ & &    & A3 & $279$ & $271$ & $331$ & $191$\\
    & A4 & $2$  & $2$   & $10$ & $45$ & &    & A4 & $99$ & $126$ & $179$ & $97$\\
    & A5 & \--- & \---  & $12$ & $19$ & &    & A5 & $36$ & $35$ & $99$ & $79$\\
\hline
\end{tabular}
\end{center}
\end{table}

\subsection{Rochdale data}

\begin{figure}
 \centering
 \makebox{\includegraphics[width=6.5in,angle=0]{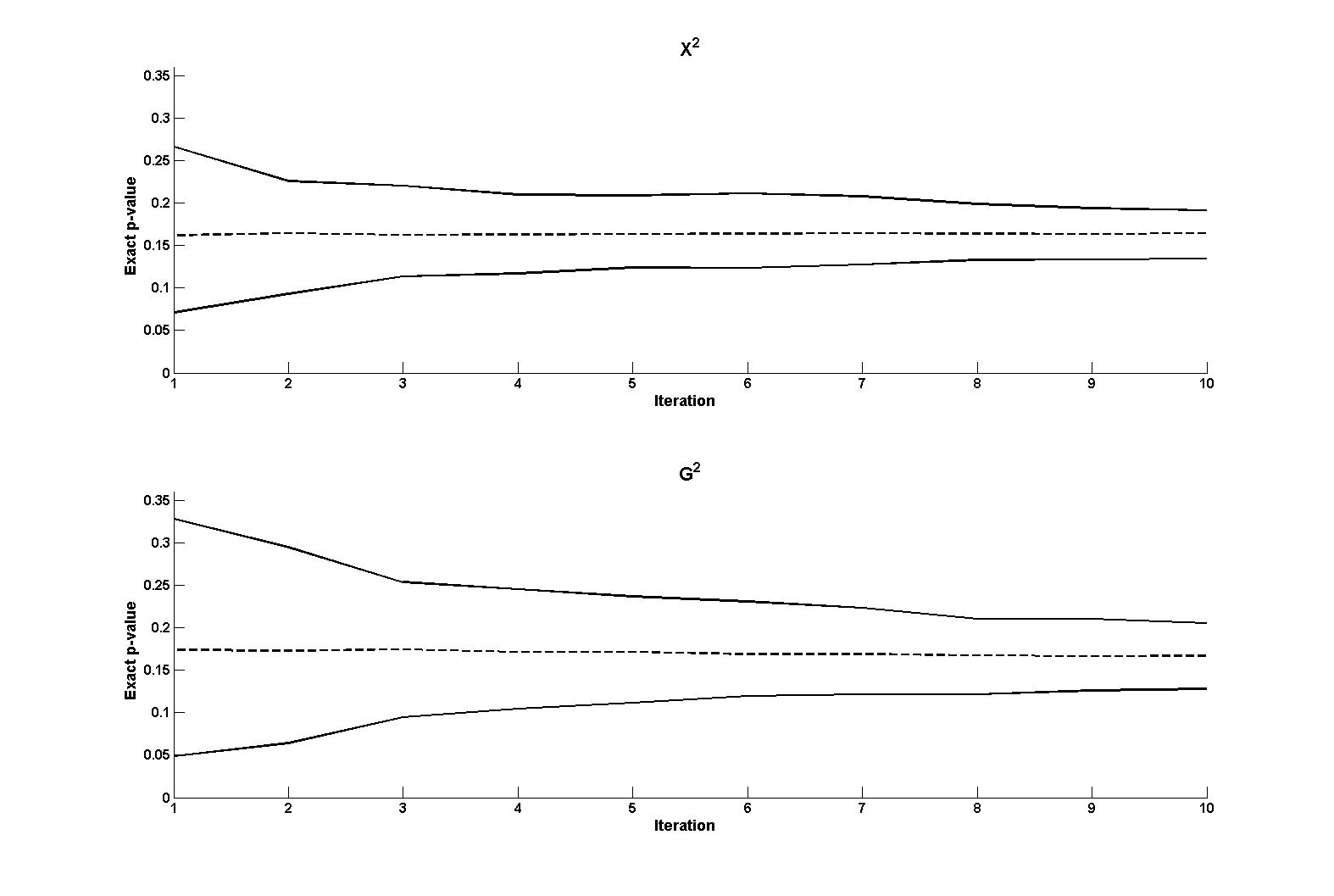}}
  \caption{\label{fig:rochdaleconvergence} Convergence of $100$ independent Markov chains based on the dynamic Markov basis for the Rochdale data and the all two-way interaction model. The upper panel shows the convergence to the estimate $0.1642$ of the $X^{2}$ exact p-value, while the lower panel shows the convergence to the estimate $0.1668$ of the $G^{2}$ exact p-value. The $x$-axis gives the number of iterations in increments of $250000$. For each chain we calculated p-value estimates based on $250000i$ sampled tables with $i=1,2,\ldots,10$. The dotted line represents the mean of these incremental estimates, while the solid lines represent their $2.5\%$ and $97.5\%$ quantiles.}
\end{figure}

\begin{figure}
 \centering
 \makebox{\includegraphics[width=5in,angle=0]{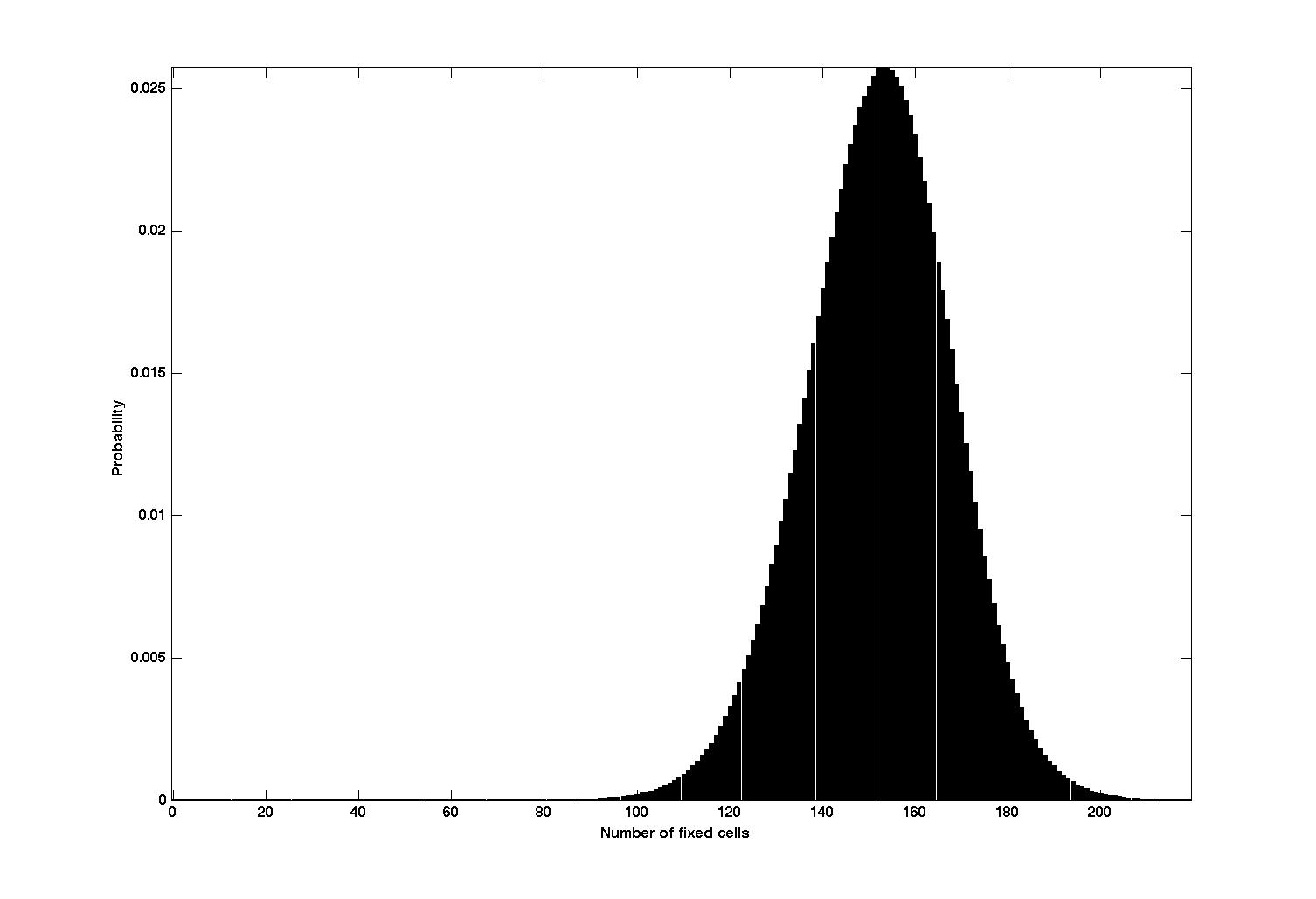}}
  \caption{\label{fig:grochdale} The discrete distribution $g_{T}(\cdot)$ for the Rochdale data and the all two-way interaction model as determined by Algorithm \ref{alg:gdot}. The number of free cells is $219$. Remark that $g_{T}(\cdot)$ is considerably more diffuse than the corresponding distribution for the NBER data \--- see Figure \ref{fig:nberdiagnostic}.}
\end{figure}

\noindent The data in Table \ref{tab:rochdaledata} is a cross-classification of eight binary variables relating women's economic activity and husband's unemployment from a survey of households in Rochdale \--- see Whittaker \shortcite{whittaker1990} page 279. The variables are as follows: $a$, wife economically active (no,yes); $b$, age of wife $>38$ (no,yes); $c$, husband unemployed (no,yes); $d$, child $\le 4$ (no,yes); $e$, wife's education, high-school+ (no,yes); $f$, husband's education, high-school+ (no,yes); $g$, Asian origin (no,yes); $h$, other household member working (no,yes). There are $665$ individuals cross-classified in $256$ cells, which means that the mean number of observations per cell is $2.6$. The table has $165$ counts of zero and $217$ other cells contain at most three observations.\\
\indent Whittaker \shortcite{whittaker1990} argues that this table is sparse and subsequently that the applicability of any asymptotic results relating to the limiting distributions of goodness-of-fit statistics for log-linear models becomes questionable due to the zeros present in marginals of dimension three or more. The likelihood-ratio test statistic for the all two-way interaction model is $G^{2}=144.59$, while  the observed $X^{2}$ test statistic is $258.65$. The all two-way interaction model has 219 degrees of freedom, which leads to asymptotic p-values of 1 for the $G^{2}$ statistic and of $0.034$ for the $X^{2}$ statistic.\\
\indent The Markov chain algorithm with dynamic Markov bases and the CB algorithm give similar estimates of the exact p-values. More specifically, the exact $G^{2}$ p-value is estimated to be $0.1668\pm 0.0684$ by our approach  and $0.1644\pm 0.0443$ by the CB approach.  The exact $X^{2}$ p-value is estimated to be $0.1642\pm 0.0524$ by our approach and $0.1717\pm 0.1101$ by the CB approach. The Monte Carlo standard errors for the $G^{2}$ p-value of both Markov chain algorithms are comparable, but the CB algorithm gives a larger standard error when computing the $X^{2}$ p-value. In a recent paper, Dinwoodie and Chen \shortcite{dinwoodie-chen-2010} report two different estimates ($0.223\pm 0.091$ and $0.186\pm 0.041$) of the exact $G^{2}$ p-value obtained with their new version of the SIS algorithm based on two cell orderings. We found estimates equal to $1$ for both the $G^{2}$ and $X^{2}$ exact p-values using our implementation of the SIS algorithm of Chen et al. \shortcite{chen-annals2006}.\\
\indent Figure \ref{fig:rochdaleconvergence} illustrates the convergence of the Markov chain algorithm from Section \ref{sec:markovchain} across its $100$ replicates. Its running time is $18821.75\pm 304.01$ seconds per $250000$ iterations. Our Markov chain algorithm makes use of an estimate of the discrete distribution $g_{T}(\cdot)$ that is obtained by running Algorithm \ref{alg:gdot} for $10$ million iterations. It takes approximately $8813.6\pm 40.62$ seconds per $100000$ sampled tables to obtain the distribition $g_{T}(\cdot)$ from Figure \ref{fig:grochdale}.

{ \renewcommand{\baselinestretch}{1.0}
\begin{table}[h]
\begin{center}
\caption{\label{tab:rochdaledata}Rochdale data from Whittaker \shortcite{whittaker1990}. The cells counts are written in lexicographical order with $h$ varying fastest and $a$ varying slowest. The grand total of this table is $665$.}
\setlength{\tabcolsep}{1mm}
\begin{tabular}{cccccccccccccccc}\hline
 $5$  & $0$ & $2$  &  $1$  & $5$  & $1$ & $0$ & $0$ & $4$ & $1$ & $0$ & $0$ & $6$ & $0$ & $2$ & $0$\\
 $8$  & $0$ & $11$ & $0$ & $13$  & $0$ & $1$ & $0$ & $3$ & $0$ & $1$ & $0$ & $26$ & $0$ & $1$ & $0$\\
 $5$  & $0$ & $2$ & $0$ & $0$ & $0$ & $0$ & $0$ & $0$ & $0$ & $0$ & $0$ & $0$ & $0$ & $1$ & $0$\\
 $4$  & $0$ & $8$ & $2$ & $6$ & $0$ & $1$ & $0$ & $1$ & $0$ & $1$ & $0$ & $0$ & $0$ & $1$ & $0$\\
 $17$ & $10$ & $1$ & $1$ & $16$ & $7$ & $0$ & $0$ & $0$ & $2$ & $0$ & $0$ & $10$ & $6$ & $0$ & $0$\\
 $1$ & $0$ & $2$ & $0$ & $0$ & $0$ & $0$ & $0$ & $1$ & $0$ & $0$ & $0$ & $0$ & $0$ & $0$ & $0$\\
 $4$ & $7$ & $3$ & $1$ & $1$ & $1$ & $2$ & $0$ & $1$ & $0$ & $0$ & $0$ & $1$  & $0$ & $0$ & $0$\\
 $0$ & $0$ & $3$ & $0$ & $0$ & $0$ & $0$ & $0$ & $0$ & $0$ & $0$ & $0$ & $0$ & $0$ & $0$ & $0$\\
 $18$ & $3$ & $2$ & $0$ & $23$ & $4$ & $0$ & $0$ & $22$ & $2$ & $0$ & $0$ & $57$ & $3$ & $0$ & $0$\\
 $5$ & $1$ & $0$ & $0$ & $11$ & $0$ & $1$ & $0$ & $11$ & $0$ & $0$ & $0$ & $29$ & $2$ & $1$ & $1$\\
 $3$ & $0$ & $0$ & $0$ & $4$ & $0$ & $0$ & $0$ & $1$ & $0$ & $0$ & $0$ & $0$ & $0$ & $0$ & $0$\\
 $1$ & $1$ & $0$ & $0$ & $0$ & $0$ & $0$ & $0$ & $0$ & $0$ & $0$ & $0$ & $0$ & $0$ & $0$ & $0$\\
 $41$ & $25$ & $0$ & $1$ & $37$ & $26$ & $0$ & $0$ & $15$ & $10$ & $0$ & $0$ & $43$ & $22$ & $0$ & $0$\\
 $0$ & $0$ & $0$ & $0$ & $2$ & $0$ & $0$ & $0$ & $0$ & $0$ & $0$ & $0$ & $3$ & $0$ & $0$ & $0$\\
 $2$ & $4$ & $0$ & $0$ & $2$ & $1$ & $0$ & $0$ & $0$ & $1$ & $0$ & $0$ & $2$ & $1$ & $0$ & $0$\\
 $0$ & $0$ & $0$ & $0$ & $0$ & $0$ & $0$ & $0$ & $0$ & $0$ & $0$ & $0$ & $0$ & $0$ & $0$ & $0$\\
\hline
\end{tabular}
\end{center}
\end{table}
}

\section{Conclusions} \label{sec:conclusions}

\noindent In this paper we introduced dynamic Markov bases and proposed a Markov chain algorithm for sampling tables based on them. Our methods are applicable off-the-shelf to calculate exact p-values for reference sets of tables defined by any type of linear and bounds constraints. The choice of distribution $g_{T}(\cdot)$ that is used in the mixture instrumental distribution (\ref{eq:proposalmixture}) is key for a successful application of the Markov chain algorithm described in Section \ref{sec:markovchain}. The running time of our sampling approach is a function of the expected number of optimization problems (\ref{eq:linprogbounds}), i.e.
\begin{eqnarray*}
Q(g_{T}) & = & 2(m_{c}-m^{\prime}_{r}-E_{g_{T}}(M)),
\end{eqnarray*}
\noindent that need to be solved to generate one candidate table from (\ref{eq:proposalmixture}). In our NBER data example, the number of free cells is $26$ which yields $Q(g_{T})=5.46$ for the distribution $g_{T}(\cdot)$ from Figure \ref{fig:nberdiagnostic}. By comparison, if we would work with the uniform distribution $g_{T}(\cdot)=\Uni_{(0:(m_{c}-m^{\prime}_{r}-1))}(\cdot)$ in the instrumental distribution (\ref{eq:proposalmixture}), the expected number of optimization problems increases to $Q(\Uni_{(0:25)})=27$. For the Rochdale data example we obtain $Q(g_{T})=134.1$ for the distribution $g_{T}(\cdot)$ from Figure \ref{fig:grochdale} and $Q(\Uni_{(0:218)})=220$. As such, Algorithm \ref{alg:gdot} is quite effective in determining distributions $g_{T}(\cdot)$ that are lead to Markov chains with dynamic Markov bases with good mixing properties and reasonable running times. Finding suitable distributions $g_{T}(\cdot)$ that are properly adapted to a reference set of tables $T$ in the absence of a well-defined procedure could be detrimental in practice, hence Algorithm \ref{alg:gdot} should be seen as integral part of the dynamic Markov bases methodology we proposed.\\
\indent We hope that the basic idea of generating only the moves needed to complete one iteration of the random walk will be adopted by other researchers since it is a more practical alternative to the determination of the entire Markov basis in one computationally intensive step as it was originally suggested in Diaconis and Sturmfels \shortcite{diaconis-sturmfels-1998}. Relevant questions relate to studying the theoretical properties of dynamic Markov bases using algebraic statistics in the spirit of Rapallo \shortcite{rapallo-2006}, Aoki and Takemura \shortcite{aoki-takemura-2010} and Rapallo and Yoshida \shortcite{rapallo-yoshida-2010}. These research directions should be added to the list of open problems related to Markov bases presented in Yoshida \shortcite{yoshida-2010}.\\

\section*{Supplemental Material}

{\bf Computer Code and Data}: Supplemental materials for this article are contained in a single
zip archive and can be obtained in a single download. This archive contains the datasets NBER and Rochdale
(in text files)  as well as the C++ source code to run the algorithms
described in this article (the Markov chain based on the dynamic Markov bases and the sequential importance sampling algorithm). A detailed description of the files contained in this archive is
contained in a README.txt file enclosed in the archive.

\section*{Acknowledgments}

This work was partially supported by a seed grant from the Center of Statistics and the Social Sciences, University of Washington. The author thanks Anna Klimova for her assistance with some of the numerical results presented in the paper. The author thanks three anonymous reviewers and the AE for their helpful comments.

\bibliographystyle{jasa}
\bibliography{dobra-jcgs-paper}

\end{document}